\documentclass[preprint]{emulateapj}

\usepackage{cancel,soul,ulem,amsmath}
\usepackage{color}
\usepackage{multirow}
\usepackage{rotating}

\def\13co{$^{13}$CO}

\def\av{$A_{V}$}
\def\c18o{C$^{18}$O}
\def\cc{cm$^{-3}$}
\def\co{$^{12}$CO}
\def\cm2{cm$^{-2}$}

\def\ebv{$E(B{-}V)$}
\def\h2{H$_2$}
\def\hi{H{\sc i}}

\def\k0{$\kappa_{353}$}

\def\nh3{NH$_3$}
\def\n2h{N$_2$H$^+$}
\def\NHI{$N_\mathrm{HI}$}
\def\nhiUnit{$\times$10$^{20}$cm$^{-2}$}
\def\NHIthin{$N^{*}_\mathrm{HI}$}
\def\NHm{$N_{\mathrm{H}_{2}}$}
\def\NH{$N_\mathrm{H}$}
\def\nh{$n_\mathrm{H}$}
\def\NOH{$N_\mathrm{OH}$}

\def\s{s$^{-1}$}
\def\s353{$\sigma_{353}$}
\def\sigUnit{$\times$10$^{-27}$cm$^{2}$H$^{-1}$}

\def\Tex{$T_{\mathrm{ex}}$}
\def\Ts{$T_{\mathrm{s}}$}
\def\t353{$\tau_{353}$}

\def\xco{$X_\mathrm{CO}$}
\def\xoh{$X_\mathrm{OH}$}

\newcommand{\Rmnum}[1]{\expandafter\@slowromancap\romannumeral #1@}
\makeatother

\shorttitle{Dust-Gas Scaling Relations}
\shortauthors{Hiep Nguyen et al. (2018)}

\begin{document}

\title{Dust-Gas Scaling Relations and OH Abundance in the Galactic ISM}

\author{Hiep Nguyen\altaffilmark{1,2}, 
J. R. Dawson\altaffilmark{1,2},
M.-A. Miville-Desch{\^e}nes\altaffilmark{3,4},
Ningyu Tang\altaffilmark{5},
Di Li\altaffilmark{5,6,7},
Carl Heiles\altaffilmark{8},
Claire E. Murray\altaffilmark{9},
Sne\v{z}ana Stanimirovi{\'c}\altaffilmark{10},
Steven J. Gibson\altaffilmark{11},
N. M. McClure-Griffiths\altaffilmark{12},
Thomas Troland\altaffilmark{13},
L. Bronfman\altaffilmark{14},
R. Finger\altaffilmark{15}
}

\altaffiltext{1}{Department of Physics and Astronomy and MQ Research Centre in Astronomy, Astrophysics and Astrophotonics, Macquarie University, NSW 2109, Australia. Email: van-hiep.nguyen@hdr.mq.edu.au}
\altaffiltext{2}{Australia Telescope National Facility, CSIRO Astronomy and Space Science, PO Box 76, Epping, NSW 1710, Australia}
\altaffiltext{3}{Institut d'Astrophysique Spatiale, CNRS, Univ. Paris-Sud, Universit{\'e} Paris-Saclay, B{\^a}t. 121, 91405, Orsay Cedex, France}
\altaffiltext{4}{Laboratoire AIM, Paris-Saclay, CEA/IRFU/DAp - CNRS - Universit{\'e} Paris Diderot, 91191, Gif-sur-Yvette Cedex, France}
\altaffiltext{5}{National Astronomical Observatories, CAS, Beijing 100012, China}
\altaffiltext{6}{Key Laboratory of Radio Astronomy, Chinese Academy of Science}
\altaffiltext{7}{University of Chinese Academy of Sciences, Beijing 100049, China}
\altaffiltext{8}{Department of Astronomy, University of California, Berkeley, 601 Campbell Hall 3411, Berkeley, CA 94720-3411}
\altaffiltext{9}{Space Telescope Science Institute, 3700 San Martin Drive, Baltimore, MD 21218, USA}
\altaffiltext{10}{Department of Astronomy, University of Wisconsin-Madison, 475 North Charter Street, Madison, WI 53706, USA}
\altaffiltext{11}{Department of Physics and Astronomy, Western Kentucky University, Bowling Green, KY 42101, USA}
\altaffiltext{12}{Research School of Astronomy and Astrophysics, Australian National University, Canberra, ACT 2611, Australia}
\altaffiltext{13}{Department  of Physics  and Astronomy,  University of
Kentucky, Lexington, Kentucky 40506}
\altaffiltext{14}{Departamento de Astronom\'ia, Universidad de Chile,
Casilla 36, Santiago de Chile, Chile}
\altaffiltext{15}{Astronomy Department, Universidad de Chile, Camino El Observatorio 1515, 1058 Santiago, Chile}

\begin{abstract}
Observations of interstellar dust are often used as a proxy for total gas column density \NH. By comparing \textit{Planck} thermal dust data (Release 1.2) and new dust reddening maps from Pan-STARRS 1 and 2MASS \citep{Green2018}, with accurate (opacity-corrected) \hi\ column densities and newly-published OH data
from the Arecibo Millennium survey and 21-SPONGE, we confirm linear correlations between dust optical depth \t353, reddening \ebv and the total proton column density \NH\ in the range (1--30)\nhiUnit, along sightlines with no molecular gas detections in emission. We derive an \NH/\ebv\ ratio of (9.4$\pm$1.6)$\times$10$^{21}$cm$^{-2}$mag$^{-1}$ for purely atomic sightlines at $|b|$$>$5$^{\circ}$, which is 60\% higher than the canonical value of \citet{Bohlin1978}. We report a $\sim$40\% increase in opacity {\s353}=\t353/\NH, when moving from the low column density (\NH$<$5\nhiUnit) to moderate column density (\NH$>$5\nhiUnit) regime, and suggest that this rise is due to the evolution of dust grains in the atomic ISM. Failure to account for \hi\ opacity can cause an additional apparent rise in \s353, of the order of a further $\sim$20\%. We estimate molecular hydrogen column densities \NHm\ from our derived linear relations, and hence derive the OH/\h2\ abundance ratio of \xoh$\sim$1$\times$10$^{-7}$ for all molecular sightlines. Our results show no evidence of systematic trends in OH abundance with \NHm\ in the range \NHm$\sim$(0.1$-$10)$\times$10$^{21}$cm$^{-2}$. This suggests that OH may be used as a reliable proxy for \h2\ in this range, which includes sightlines with both CO-dark and CO-bright gas.
\end{abstract}


\section{Introduction}
Observations of neutral hydrogen in the interstellar medium (ISM) have historically been dominated by two radio spectral lines: the 21 cm line of atomic hydrogen (\hi) and the microwave emission from carbon monoxide (CO), particularly the CO(J=1--0) line. The former provides direct measurements of the warm neutral medium (WNM), and the cold neutral medium (CNM) which is the precursor to molecular clouds. The latter is widely used as a proxy for molecular hydrogen (\h2), often via the use of an empirical `X-factor', (e.g. \citealt{Bolatto2013}). The processes by which CNM and molecular clouds form from warm atomic gas sows the seeds of structure into clouds, laying the foundations for star formation. Being able to observationally track the ISM through this transition is of key importance. 
\par
However, there is strong evidence for gas not seen in either \hi\ or CO. This undetected material is often called ``dark gas'', following \cite{Grenier2005}. These authors found an excess of diffuse gamma-ray emission from the Local ISM, with respect to the expected flux due to cosmic ray interactions with the gas mass estimated from \hi\ and CO. 
Similar conclusions have been reached using many different tracers, including $\gamma$-rays (e.g. \citealt{Abdo2010}, \citealt{Ackermann2012}, \citealt{Ackermann2011}), infra-red emission from dust (e.g. \citealt{Blitz1990, Reach1994, Douglas2007, PLC2011, PLC2014}), dust extinction \citep[e.g.][]{Paradis2012, Lee2015}, C{\sc ii} emission (\citealt{Pineda2013, Langer2014, Tang2016}) and OH 18 cm emission and absorption (e.g. \citealt{Wannier1993, Liszt1996, Barriault2010, Allen2012, Allen2015, Engelke2018}).

While a minority of studies have suggested that cold, optically thick \hi\ could account for almost all the missing gas mass \citep{Fukui2015}, CO-dark \h2 is generally expected to be a major constituent, particularly in the envelopes of molecular clouds \citep[e.g.][]{Lee2015}. In diffuse molecular regions, \h2 is effectively self-shielded, but CO is typically photodissociated \citep{Tielens1985a,Tielens1985b,vanDishoeck1988,Wolfire2010,Glover2011,Lee2015,Glover2016}, meaning that CO lines are a poor tracer of \h2 in such environments. Indeed \textit{Herschel} observations of C{\sc ii} suggest that between 20--75\% of the \h2 in the Galactic Plane may be CO-dark \citep[][]{Pineda2013}.

For the atomic medium, the mass of warm \hi\ can be computed directly from measured line intensities under the optically thin assumption. However, cold \hi\ with spin temperature $T_{s}$ $\lesssim$ 100K suffers from significant optical depth effects, leading to an underestimation of the total column density. This difficulty is generally addressed by combining \hi\ absorption and emission profiles observed towards (and immediately adjacent to) bright, compact continuum background sources. Such studies find that the optically thin assumption underestimates the true \hi\ column by no more than a few 10\% along most Milky Way sightlines (e.g. \citealt{Dickey1983,Dickey2000,Dickey2003,Heiles2003a,Heiles2003b,Liszt2014,Lee2015}), although the fraction missed in some localised regions may be much higher \citep{Bihr2015}. 

Since dust and gas are generally well mixed, absorption due to dust grains has been widely used as a proxy for total gas column density. Early work (e.g. \citealt{Savage1972}, \citealt{Bohlin1978}) observed Lyman-$\alpha$ and \h2\ absorption in stellar spectra to calibrate the relationship between total Hydrogen column density \NH, and the color excess \ebv. Similar work was carried out by comparing X-ray absorption with optical extinction, \av\ (\citealt{Reina1973}, \citealt{Gorenstein1975}). Bohlin et al's value of \NH/\ebv=5.8$\times$10$^{21}$cm$^{-2}$mag$^{-1}$ has become a widely accepted standard. 

Dust emission is also a powerful tool, and requires no background source population. The dust emission spectrum in the bulk of the ISM peaks in the FIR-to-millimeter range, and arises mostly from large grains in thermal equilibrium with the ambient local radiation field (\citealt{Draine2003}, \citealt{Draine2007}). It has long been recognized that FIR dust emission could potentially be a better tracer of \NH\ than \hi\ and CO (de Vries et al.1987, Heiles et al.1988, Blitz et al.1990, Reach et al. 1994). An excess of dust intensity and/or optical depth above a linear correlation with \NH\ (as measured by \hi\ and CO) is typically found in the range \av=0.3$-$2.7 mag (\citealt{PLC2011,PLC2014,PlanckInt2014,Martin2012}), consistent with the range where CO-dark \h2\ can exist. Alternative explanations cannot be definitively ruled out, however. These include (1) the evolution of dust grains across the gas phases; (2) underestimation of the total gas column due to significant cold \hi\ opacity; (3) insufficient sensitivity for CO detection. It has also been impossible to rule out remaining systematic effects in the \textit{Planck} data or bias in the estimate of \t353\ introduced by the choice of the modified black-body model.

In this study, we examine the correlations between accurately-derived \hi\ column densities and dust-based proxies for $N_\mathrm{H}$. We make use of opacity-corrected \hi\ column densities derived from two surveys: the Arecibo Millennium Survey \citep[MS,][hereafter HT03]{Heiles2003b}, and 21-SPONGE (\citealt{Murray2015}), both of which used on-/off-source measurements towards extragalactic radio continuum sources to derive accurate physical properties for the atomic ISM. We also make use of archival OH data from the Millennium Survey, recently published for the first time in a companion paper, \citet{Li2018}. OH is an effective tracer of diffuse molecular regions (\citealt{Wannier1993, Liszt1996, Barriault2010, Allen2012, Allen2015, Xu2016, Li2018}), and has recently been surveyed at high sensitivity in parts of the Galactic Plane \citep{Dawson2014, Bihr2015}. There exists both theoretical and observational evidence for the close coexistence of interstellar OH and \h2. Observationally, they appear to reside in the same environments, as evidenced by tight relations between their column densities \citep{Weselak2014}. Theoretically, the synthesis of OH is driven by the ions O$^{+}$ and H$_3^+$ but requires \h2\ as the precursor; once \h2\ becomes available, OH can be formed efficiently through the charge-exchange chemical reactions initiated by cosmic ray ionization \citep{vanDishoeck1986}. Here we combine \hi, OH and dust datasets to obtain new measurements of the abundance ratio, \xoh=\NOH/\NHm\ -- a key quantity for the interpretation of OH datasets.

The structure of this article is as follows. In Section \ref{sec:observations}, the observations, the data processing techniques and corrections on \hi\ are briefly summarized. In Section \ref{sec:OH_data_analysis}, the results from OH observations are discussed. Section \ref{sec:proxies-for-nh} discusses the relationship between \t353, \ebv\ and \NH\ in the atomic ISM. We finally estimate the OH/\h2\ abundance ratio in Section \ref{sec:xoh}, before concluding in Section \ref{sec:conclusion}.

\section{Datasets}
In this study, we use the all-sky optical depth (\t353) map of the dust model data measured by Planck/IRAS (\citealt{PLC2014} -- hereafter PLC2014a), the reddening \ebv\ all-sky map from \citet{Green2018}, \hi\ data from both the 21-SPONGE Survey (\citealt{Murray2015}) and the Millennium Survey (\citealt{Heiles2003a}, HT03), OH data from the Millennium Survey \citep{Li2018}, CO data from the Delingha 14m Telescope, the Caltech Submillimeter Observatory (CSO) and the IRAM 30m telescope \citep{Li2018}. 
\label{sec:observations}

\subsection{\hi\ and OH}
\label{subsec:data_hi_oh}
\hi\ data from the Millennium Arecibo 21-Centimeter Absorption-Line Survey (hereafter MS) was taken towards 79 strong radio sources (typically S $\gtrsim$ 2 Jy) using the Arecibo L-wide receiver.
The two main lines of ground state OH at 1665.402 and 1667.359 MHz were observed simultaneously towards 72 positions, and OH absorption was detected along 19 of these sightlines \citep[see also][]{Li2018}. The observations are described in detail by HT03. Briefly, their so-called Z16 observation pattern consists of one on-source absorption spectrum towards the background radio source and 16 off-source emission spectra with the innermost positions at 1.0 HPBW and the outermost positions at $\sqrt{2}$ HPBW from the central source. The off-source ``expected'' emission spectrum, the emission profile we would observe in the absence of the continuum source, is then estimated by modeling the 17-point measurements. In this work we use the published values of HT03 for the total \hi\ column density, \NHI\ (scaled as described below), and use the off-source (expected) MS emission profiles to compute the \hi\ column density under the optically thin assumption, \NHIthin, where required. We compute OH column densities ourselves, as described in Section \ref{sec:OH_data_analysis}. All OH emission and absorption spectra are scaled to a main-beam temperature scale using a beam efficiency of $\eta_{b}$=0.5 \citep{heiles2001}, appropriate if the OH is not extended compared to the Arecibo beam size of $3'$.

\begin{figure*}
 \center
  \includegraphics[width=\textwidth]{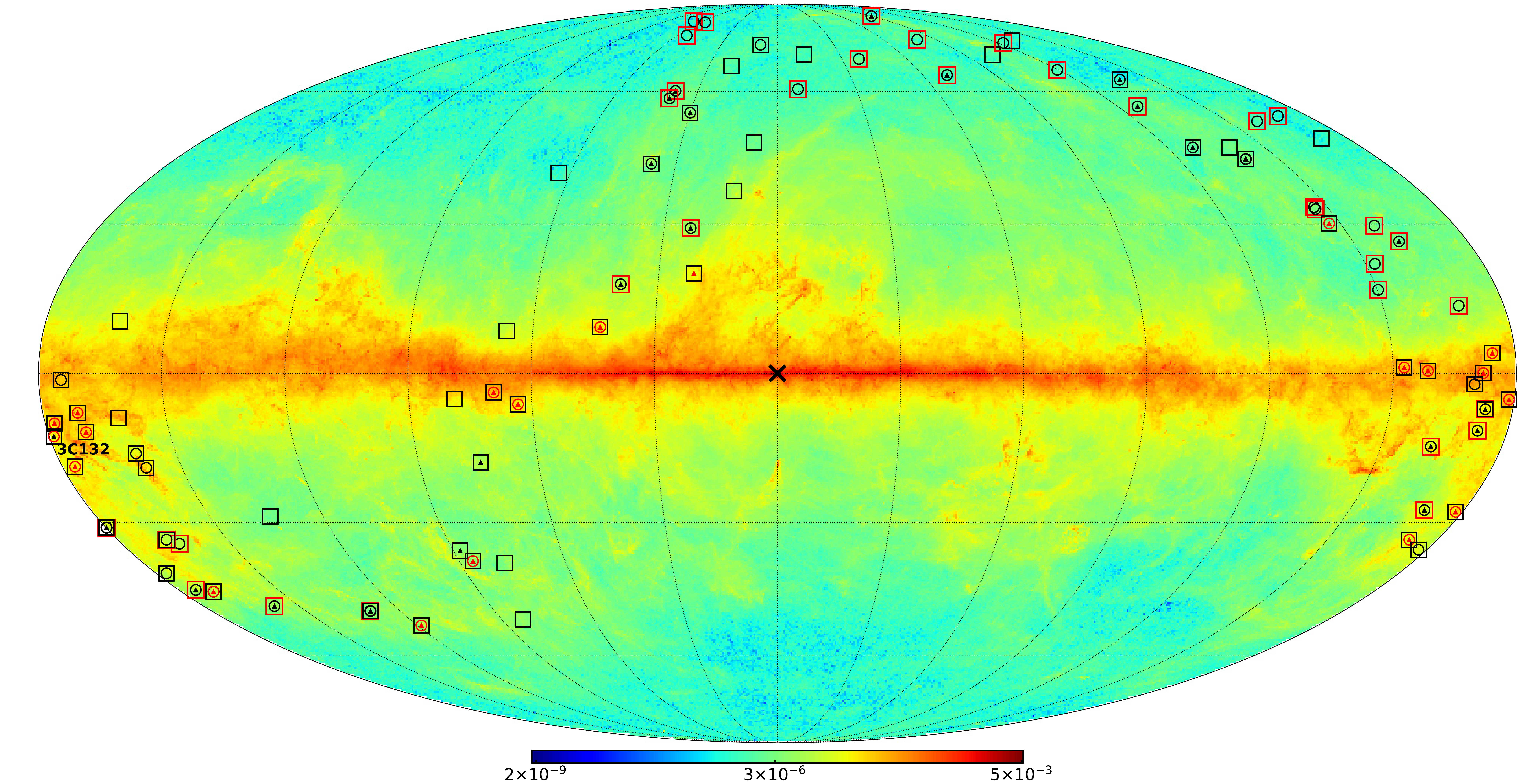}
  \caption{Locations of all 93 sightlines considered in this study, overlaid on the map of dust optical depth \t353. Squares show H{\sc i} absorption detections (93/93); red circles show OH absorption detections (19/72) and black circles non-detections (51/72); red triangles show CO detections (19/44) and black triangles non-detections (25/44). For purely atomic sightlines (those with no molecular detection at the threshold discussed in Section \ref{sec:proxies-for-nh}), the squares are colored red. Note that the absence of a symbol indicates that the sightline was not observed in that particular tracer. The labeled sightline towards 3C132 (far left) shows the single position detected in \hi\ and OH but not detected in CO emissions. The ``X'' marker labels the center of the Milky Way. Note that the symbols for a small number of sightlines entirely overlap due to their proximity on the sky.}\vspace{0.4cm}
  \label{fig:src_locations}
\end{figure*}

In order to increase the source sample, we also use \hi\ data from the Very Large Array (VLA) 21-SPONGE Survey, which observed 30 continuum sources, including 16 in-common with the Millennium Survey sample (\citealt{Murray2015}). 21-SPONGE used on-source absorption data from the VLA, combining them with off-source emission profiles observed with Arecibo. \citet{Murray2015} report an excellent agreement between the optical depths measured by the two surveys, demonstrating that the single dish Arecibo absorption profiles are not significantly contaminated with resolved 21 cm emission. Note that in this work we have used an updated scaling of the 21-SPONGE emission profiles, which applies a beam efficiency factor of 0.94 to the Arecibo spectra. The total number of unique sightlines presented in this work is therefore 93. The locations of all observed sources in Galactic coordinates are presented in Figure \ref{fig:src_locations}. Where sources were observed in both the MS and 21-SPONGE, we use the MS data.

\subsubsection{\hi\ Intensity Scale Corrections}

We check our \NHIthin\ against the Leiden-Argentine-Bonn survey \citep[LAB,][]{Hartmann1997,Kalberla2005} and the HI4PI survey \citep{hi4pi2016}. Both are widely regarded as a gold standard in the absolute calibration of Galactic \hi. We find that the optically thin column densities derived from 21-SPONGE are consistent with  LAB and HI4PI. However, the MS values are systematically lower than both LAB and HI4PI by a factor of $\sim$1.14. A possible explanation for this difference lies in the fact that (in contrast to 21-SPONGE) the MS did not apply a main beam efficiency.

To bring the MS dataset in-line with LAB, HI4PI and 21-SPONGE, one might assume that both the on-source and off-source spectra must be rescaled, and the opacity-corrected column densities recomputed according to the method of HT03 (or equivalent). However, \NHI\ may in fact be obtained from the tabulated values of HT03, with no need to perform a full reanalysis of the data. For warm components, the tabulated values of \NHI\ are simply scaled by 1.14 -- appropriate since these were originally computed directly from the the integrated off-source (expected) profiles under the optically thin assumption. For cold components, we recall that the radiative transfer equations for the on-source and off-source (expected) spectra in the MS dataset are given by:
\begin{equation}
T_\mathrm{B}^\mathrm{ON}(v) =(T_\mathrm{bg}+T_\mathrm{c})e^{-\tau_{v}} + T_\mathrm{s}(1-e^{-\tau_{v}}) + T_\mathrm{rx}
\label{eq_ton_hi}
\end{equation}
\begin{equation}
T_\mathrm{B}^\mathrm{OFF}(v)=T_\mathrm{bg}e^{-\tau_{v}} + T_\mathrm{s}(1-e^{-\tau_{v}}) + T_\mathrm{rx},
\label{eq_toff_hi} 
\end{equation}
\noindent where $T_\mathrm{B}^\mathrm{OFF} (v)$ and $T_\mathrm{B}^\mathrm{ON} (v)$ are the main beam temperatures of the off-source spectrum and on-source spectrum, respectively. $T_\mathrm{s}$ is the spin temperature, $\tau_{v}$ is the optical depth, $T_\mathrm{rx}$ is the receiver temperature ($\sim$25 K), and $T_\mathrm{c}$ is the main-beam temperature of the continuum source, obtained from the line-free portions of the on-source spectrum. $T_\mathrm{bg}$ is the continuum background brightness temperature including the 2.7 K isotropic radiation from CMB and the Galactic synchrotron background at the source position. Equations (\ref{eq_ton_hi}) and (\ref{eq_toff_hi}) may be solved for $\tau_{v}$ and $T_\mathrm{s}$:
\begin{equation}
e^{-\tau_{v}} = \frac{T_\mathrm{B}^\mathrm{ON}(v) - T_\mathrm{B}^\mathrm{OFF}(v)}{T_\mathrm{c}},
\label{eq_etau_hi} 
\end{equation}
\begin{equation}
T_\mathrm{s} = \frac{T_\mathrm{B}^\mathrm{OFF}(v)-T_\mathrm{bg}e^{-\tau_{v}} - T_\mathrm{rx}}{1-e^{-\tau_{v}}}.
\label{eq_ts_hi}
\end{equation}
From Equation (\ref{eq_etau_hi}), it is clear that optical depth is unchanged by any rescaling, which will affect both the numerator and denominator of the expression identically. Only $T_{s}$ must be recomputed. This is done on a component-by-component basis from the tabulated Gaussian fit parameters for peak optical depth, $\tau_0$, peak brightness temperature (scaled by 1.14), and the linewidth $\Delta v$. The corrected \NHI\ is obtained from
\begin{equation}
\frac{N_{\mathrm{HI}}}{[10^{18}~\mathrm{cm}^{-2}]}=1.94 \cdot \tau_0 \cdot \frac{T_{\mathrm{s}}}{[K]} \cdot \frac{\Delta v}{[\mathrm{km~s}^{-1}]},
\label{eq_nhi}
\end{equation}
\noindent where the factor 1.94 includes the usual constant 1.8224 and the 1.065 arising from the integration over the Gaussian line profile.

\subsubsection{\NHI\ vs \NHIthin}

\begin{figure}
\includegraphics[width=1.0\linewidth]{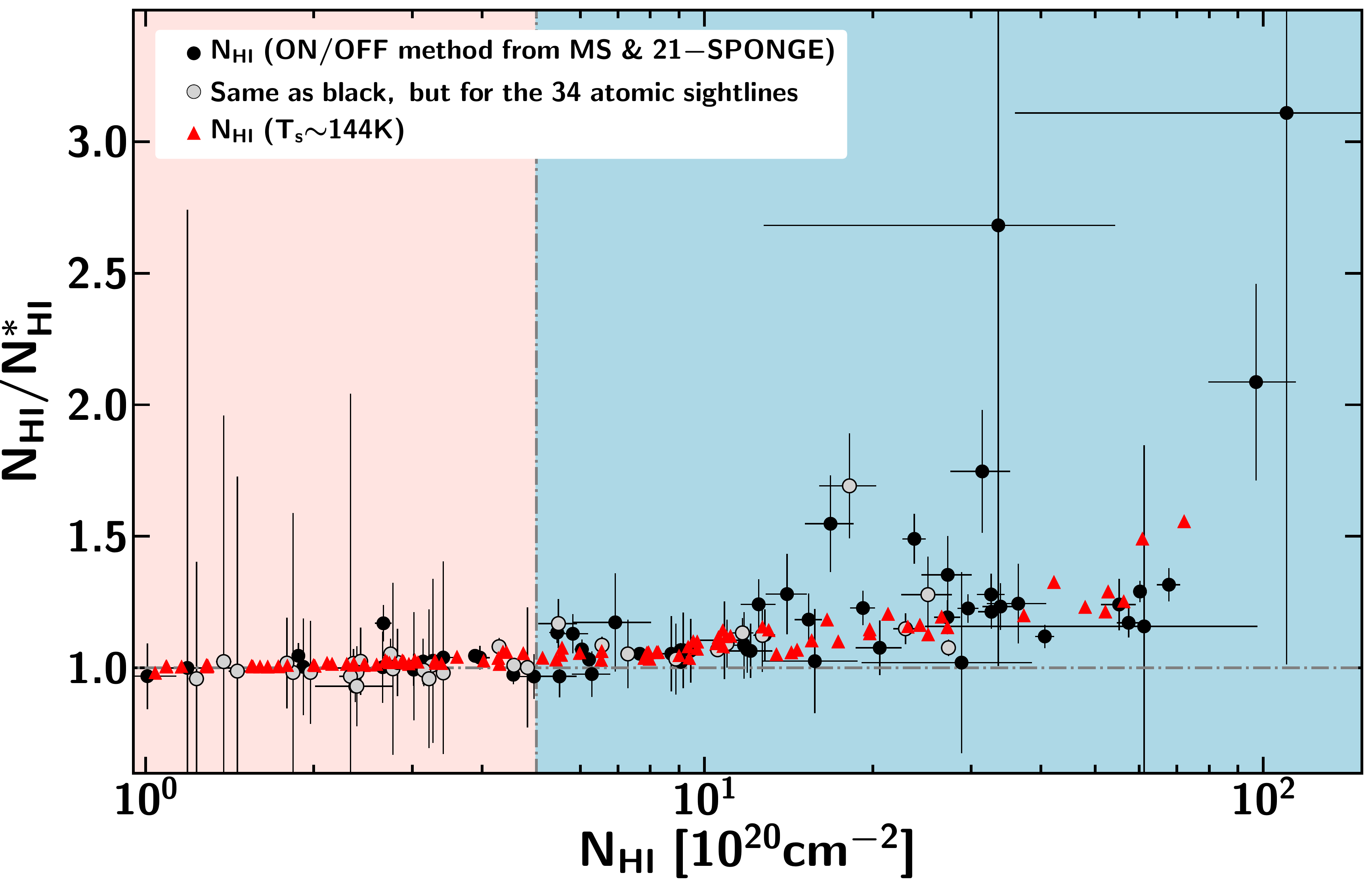}
\caption{The ratio $f$=\NHI/\NHIthin\ as a function of opacity-corrected \NHI\ along 93 sightlines from the MS and 21-SPONGE surveys. Circles show accurate \NHI\ obtained via on- and off-source observations (HT03; scaled as described in the text), with the 34 atomic sightlines (selection criteria described in Section \ref{sec:proxies-for-nh}) filled grey and all other points filled black. Red triangles show \NHI\ obtained from \NHIthin\ assuming a single isothermal component of \Ts$\sim$144 K. The vertical dashed line is plotted at \NHI=5\nhiUnit, the horizontal dashed line marks where \NHI=\NHIthin.} 
\label{fig:ratio_vs_nhi}
\end{figure}

We show in Figure \ref{fig:ratio_vs_nhi} the correlation between \NHI\ and \NHIthin\ towards all 93 positions. While optically thin \hi\ column density is comparable with the true column density in diffuse/low-density regions with \NHI$\lesssim$5\nhiUnit, opacity effects start to become apparent above $\sim$5\nhiUnit.

If a linear fit is performed to the data, the ratio $f$=\NHI/\NHIthin\ may be described as a function of log(\NHIthin/10$^{20}$) with a slope of (0.19$\pm$0.02) and an intercept of (0.89$\pm$0.02) \citep[see also][]{Lee2015}. Alternatively, a simple isothermal correction to the optically thin \NHIthin\ data with \Ts$\sim$144 K also yields a good agreement with our datapoints, as illustrated in Figure \ref{fig:ratio_vs_nhi} \citep[see also][]{Liszt2014b}. This approach also better fits the low \NHI\ plateau, \NHI$<$5\nhiUnit, below which \NHIthin$\approx$\NHI. While a single component with a constant spin temperature is a poor physical description of interstellar \hi, it can provide a reasonable (if crude) correction for opacity.

\subsection{CO}
\label{subsec:CO}
As described in \citet{Li2018}, a CO follow-up survey was conducted towards 44 of the sightlines considered in this work. The J=1--0 transitions of \co, \13co, and \c18o\ were observed with the Delingha 13.7m telescope in China. \co(J=2--1) data for 45 sources and J=3--2 data for 8 sources with strong \co\ emission were taken with the 10.4m CSO on Mauna Kea, with further supplementary data obtained by the IRAM 30m telescope. In this work we use CO data solely to identify and exclude from some parts of the analysis positions with detected CO-bright molecular gas -- comprising 19 of the 44 observed positions. These positions are identified in Figure \ref{fig:src_locations}. 

\subsection{Dust}
\label{subsec:data_planck_iras}
To trace the total gas column density \NH\ we use publicly-available all-sky maps of the 353 GHz dust optical depth (\t353) from the \textit{Planck} satellite. The \t353\ map was obtained by a modified black-body (MBB) fit to the first 15 months of 353, 545, and 857 GHz data, together with IRAS 100 micron data (for details see PLC2014a). The angular resolution of this dataset is 5 arcmin. In this work we use the R1.20 data release in Healpix\footnote{$http://healpix.sourceforge.net$} format (\citealt{Gorski2005}). 
For dust reddening, we employ the newly-released all-sky 3D dust map of \citet{Green2018} at an angular resolution of 3$'$.4-13$'$.7, which was derived from 2MASS and the latest Pan-STARRS 1 data photometry. In contrast to emission-based dust maps that depend on the modeling of the temperature, optical depth, and the shape of the emission spectrum, in maps based on stellar photometry reddening values are more directly measured and not contaminated from zodiacal light or large-scale structure. Here we convert the \citet{Green2018} \textit{Bayestar17} dust map to \ebv\ by applying a scaling factor of 0.884, as described in the documentation accompanying the data release\footnote{http://argonaut.skymaps.info/usage}.

\section{OH Data Analysis}
\label{sec:OH_data_analysis}

The Millennium Survey OH data consists of on-source and off-source `expected' spectra for each of the OH lines. In our companion paper \citet{Li2018}, we use the method of HT03 to derive OH optical depths, excitation temperatures and column densities. Namely, we obtain solutions for the excitation temperature, $T_\mathrm{ex}$, and $\tau$ via Gaussian fitting (to both the on-source and off-source spectra) that explicitly includes the appropriate treatment of the radiative transfer. In the present work we use a simpler channel-by-channel method for the derivation of $T_\mathrm{ex}$. 

The radiative transfer equations for the on-source and off-source (expected) spectra are identical to those for \hi, given above in Equations (\ref{eq_ton_hi}) and (\ref{eq_toff_hi}). All terms and their meanings are identical, with the exception that the spin temperature, $T_\mathrm{s}$ is replaced by $T_\mathrm{ex}$. $T_\mathrm{bg}$ is the continuum background brightness temperature including the 2.7 K isotropic radiation from CMB and the Galactic synchrotron background at the source position. 
For consistency with HT03 and \citet{Li2018}, we estimate the synchrotron contribution at 1665.402 MHz and 1667.359 MHz from the 408 MHz continuum map of \citet{Haslam1982}, by adopting a temperature spectral index of 2.8, such that
\begin{equation}
T_\mathrm{bg} = 2.7 + T_\mathrm{bg,408}(\nu_\mathrm{OH}/408)^{-2.8},
\label{eq4c} 
\end{equation}
\noindent resulting in typical values of around 3.5 K. The  background continuum contribution from Galactic HII regions may be safely ignored, since the continuum sources we observed are either at high Galactic latitudes or Galactic Anti-Center longitudes. Thus, in line-free portions of the off-source spectra:
\begin{equation}
T_\mathrm{B}^{OFF}(v)= T_\mathrm{bg} + T_\mathrm{rx}.
\label{eq_toff_offline}
\end{equation}
In the absence of information about the true gas distribution, we assume that OH clouds cover fully both the continuum source and the main beam of the telescope. We may therefore solve equations (\ref{eq_ton_hi}) and (\ref{eq_toff_hi}) to derive $T_\mathrm{ex}$ and $\tau_{v}$ for each of the OH lines, as shown in Equations (\ref{eq_etau_hi}) and (\ref{eq_ts_hi}) for the case of \hi.

\begin{figure}
\includegraphics[width=1.0\linewidth]{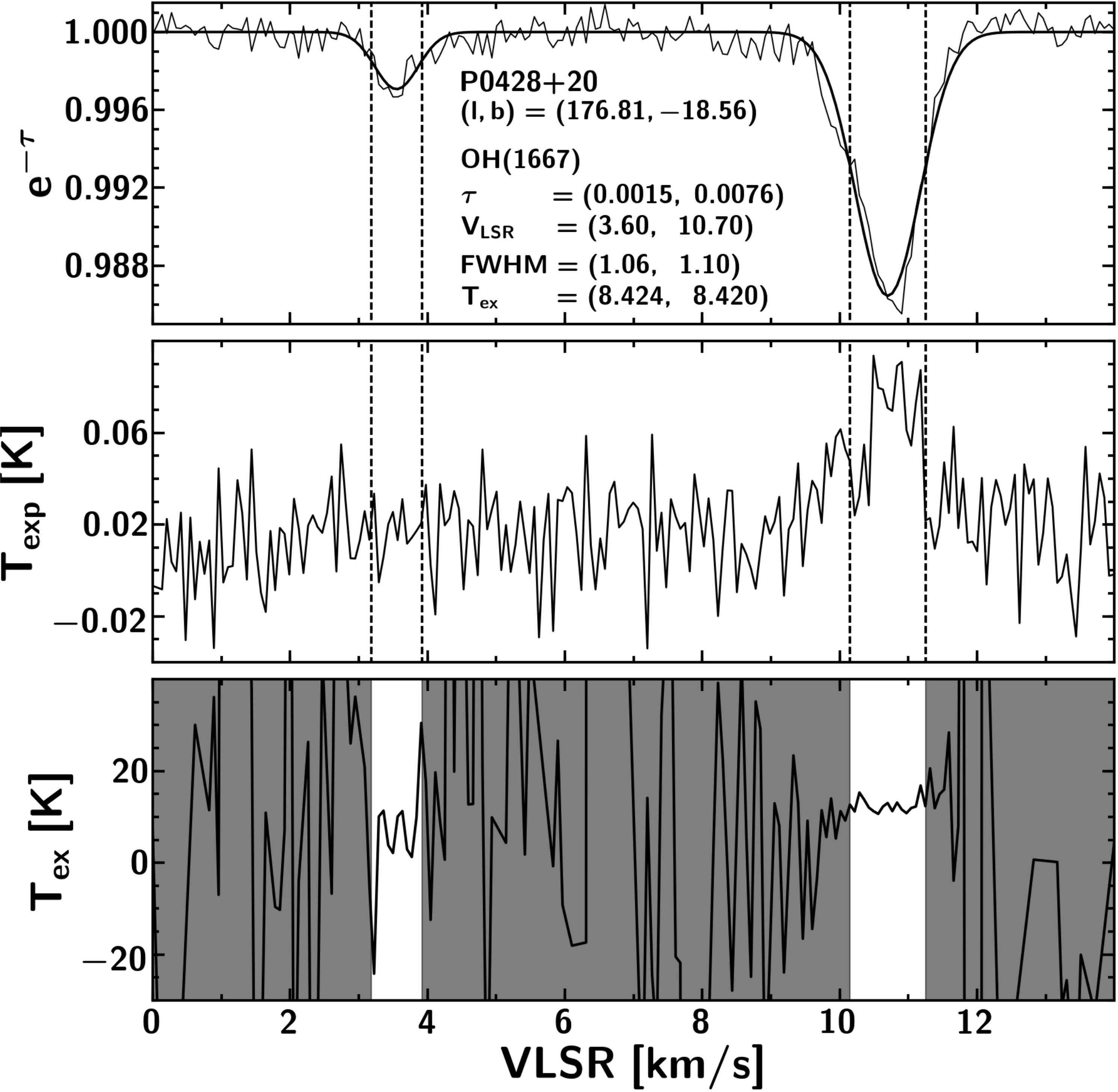}
\caption{Example of OH 1667 MHz $e^{-\tau_{v}}$ (top), expected $T_\mathrm{B}^\mathrm{OFF}$ (middle), and $T_\mathrm{ex}$ (bottom) spectra for the source P0428+20. The FWHM of the Gaussian fits to the absorption profile are used to define the range over which $T_\mathrm{ex}$ is computed for each component, shown as white regions in the bottom panel.}
\label{fig:example_p042820}
\end{figure}

We fit each OH opacity spectrum (cf. Equation \ref{eq_etau_hi}) with a set of Gaussian profiles to obtain the peak optical depth ($\tau_{0,n}$), central velocity ($v_{0,n}$) and FWHM ($\Delta v_{n}$) of each component, $n$. Equation (\ref{eq_ts_hi}) is then used to calculate excitation temperature spectra. Examples of $e^{-\tau_{v}}$, $T_\mathrm{B}^\mathrm{OFF}$, and $T_\mathrm{ex}$ spectra are shown in Figure \ref{fig:example_p042820}, together with their associated Gaussian fits. It can be seen that the $T_\mathrm{ex}$ spectra are approximately flat within the FWHM of each Gaussian component. We therefore compute an excitation temperature for each component from the mean $T_\mathrm{ex}$ in the range $v_{0,n} \pm \Delta v/2$.

\begin{figure}
\includegraphics[width=1.0\linewidth]{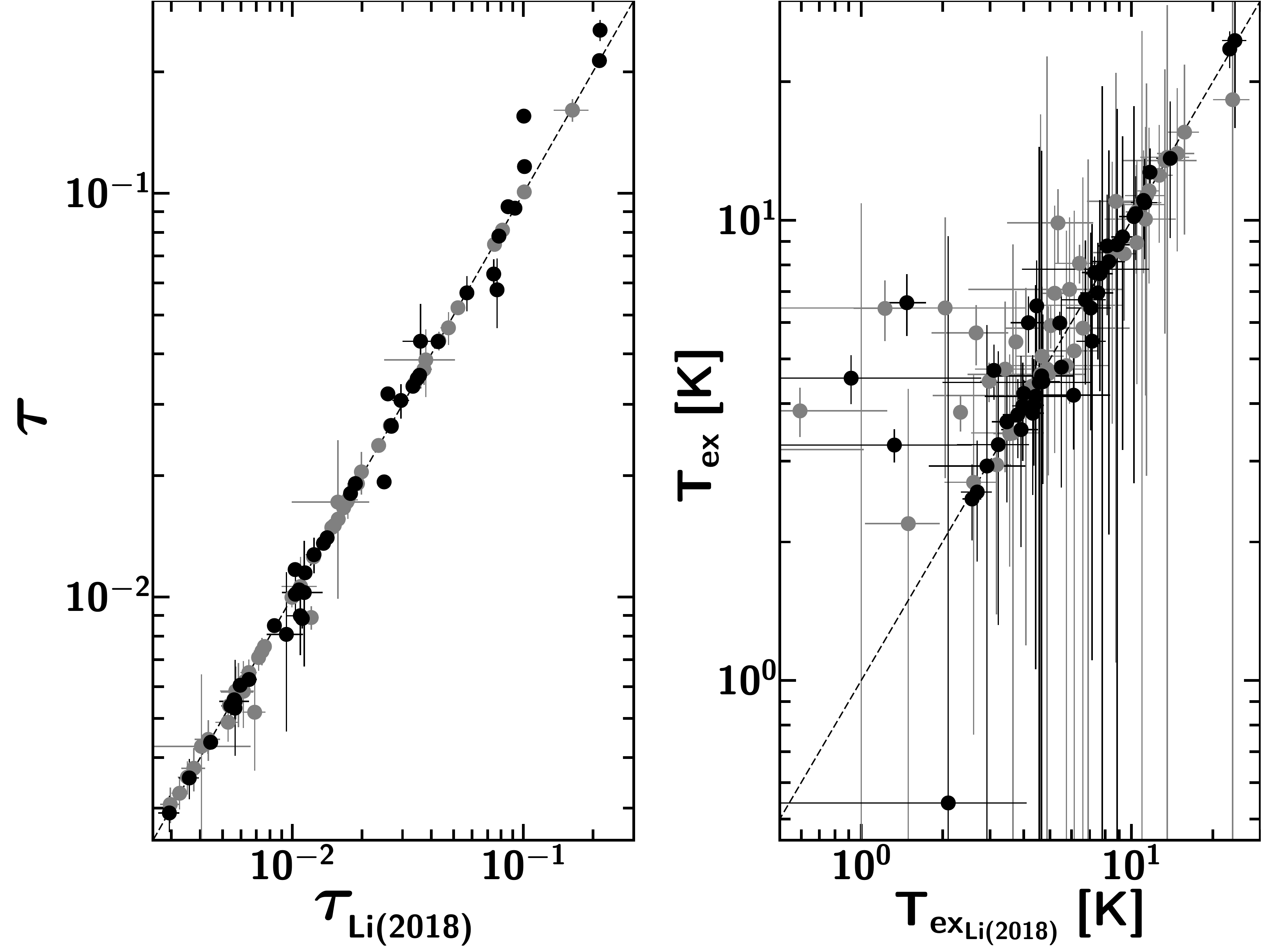}
\caption{Comparison between derived values of the peak optical depth $\tau$ (left panel), and \Tex\ (right panel) for both OH main lines, 1667 MHz (black) and 1665 MHz (gray), as obtained from our companion paper by \citet{Li2018} and the present work. The dashed lines mark where the two values are equal.}
\label{fig:tex_tau_compare}
\end{figure}

Figure \ref{fig:tex_tau_compare} compares the $\tau_0$ and \Tex\ values obtained from our method with those of \citet{Li2018}, demonstrating that the two methods generally return consistent results. Minor differences arise only for the most complex sightlines through the Galactic Plane (G197.0+1.1, T0629+10) where the spectra are not simple to analyze; however, even these points are mostly consistent to within the errors.

We compute total OH column densities, \NOH, independently from both the 1667 and 1665 MHz lines via:
\begin{equation}
\frac{N_{\mathrm{OH,1667}}}{[10^{14}~\mathrm{cm}^{-2}]}=2.39 \cdot \tau_{1667} \cdot \frac{T_\mathrm{ex,1667}}{[K]} \cdot \frac{\Delta v_{1667}}{[\mathrm{km~s}^{-2}]},
\label{eq_noh67rd}
\end{equation}
\begin{equation}
\frac{N_{\mathrm{OH,1665}}}{[10^{14}~\mathrm{cm}^{-2}]}=4.30 \cdot \tau_{1665} \cdot \frac{T_{\mathrm{ex},1665}}{[K]} \cdot \frac{\Delta v_{1665}}{[\mathrm{km~s}^{-2}]},
\label{eq_noh65rd}
\end{equation}

\noindent where the constants include Einstein A-coefficients of $A_{1667}=7.778\times 10^{-11} s^{-1}$ and $A_{1665}=7.177\times 10^{-11} s^{-1}$ for the OH main lines (\citealt{Destombes1977}). All values of $\tau_0$, $T_\mathrm{ex}$ and $N_\mathrm{OH}$ are tabulated in Table \ref{table:OH-parameters}.\\

\begin{figure*}
 \center
  \includegraphics[width=\textwidth]{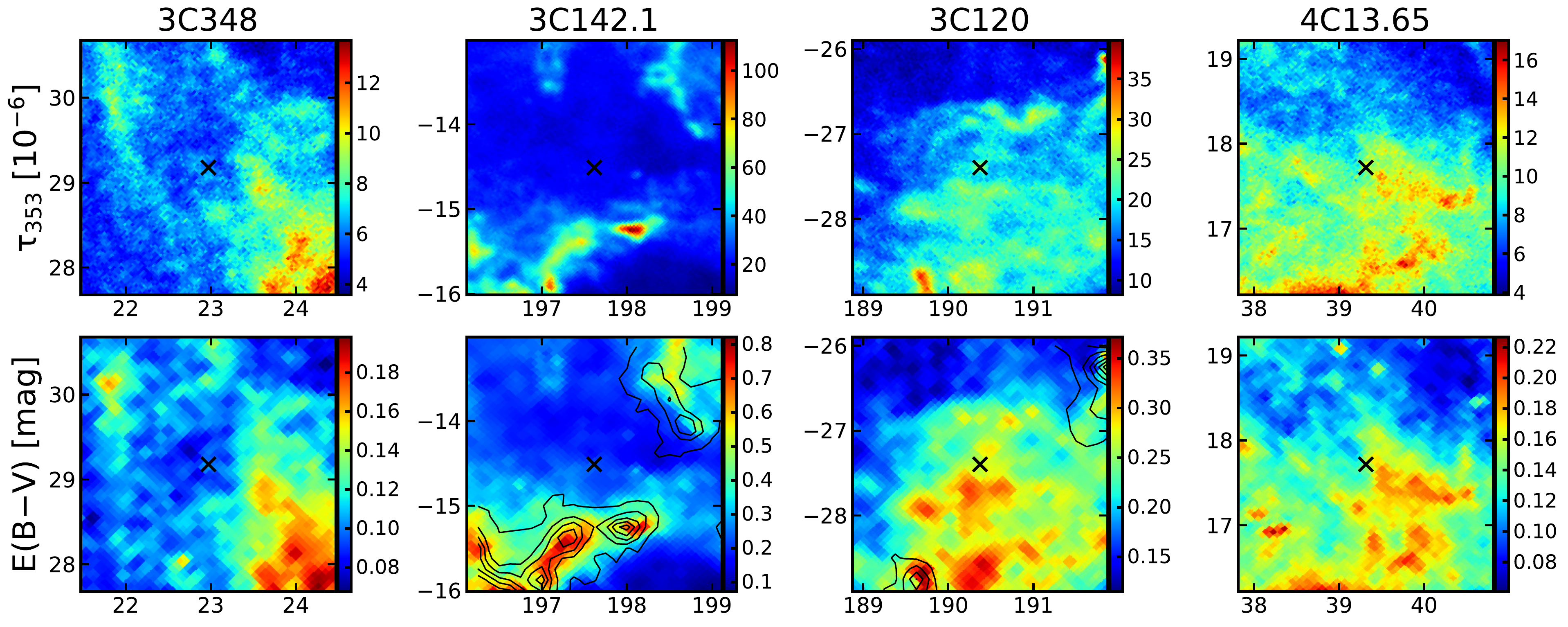}
\caption{Example maps of the immediate vicinity of 4 ``purely atomic'' sightlines towards background radio sources. Dust maps (3$\times$3 degrees in Galactic coordinates) are adopted from Planck collaboration 2014 (\t353, Nside = 2048) and \citealt{Green2018} (\ebv, Nside = 1024). The ``X'' markers show the locations of the radio sources. The contours represent the integrated intensity W$_{\mathrm{CO}(1-0)}$ from the all-sky extension to the maps of \citet{Dame2001} (T. Dame, private communication). The base level is at 0.25 K kms$^{-1}$, the typical sensitivity of CfA CO survey, and the other contour levels are evenly spaced from the base to the maximum in each map area. Equivalent maps for the remaining 30 purely atomic sightlines are given in Appendix \ref{apdxa:atomic_sightlines} and 19 OH-bright sightlines in Appendix \ref{apdx:OH_sightlines}.}
  \label{fig:src_examples}
\end{figure*}

\section{Dust-Based Proxies for Total neutral gas\\ column density}
\label{sec:proxies-for-nh}
In this section we will investigate the correlations between dust properties and the total gas column density \NH. Specifically, we consider dust optical depth at 353 GHz, \t353, and reddening, \ebv, with datasets sourced as described in Section \ref{subsec:data_planck_iras}. When these quantities are used as proxies for \NH, a single linear relationship between the measured quantity and \NH\ is typically assumed. In this work, our \hi\ dataset provides accurate (opacity-corrected) atomic column densities, while complementary OH and CO data allow us to identify and exclude sightlines with molecular gas (dark or not). We are therefore able to measure \t353/\NH\ and \ebv/\NH\ along a sample of purely atomic sightlines for which \NH\ is very well constrained.

In the following, we consider 34/93 sightlines to be ``purely atomic''. These are defined as either (a) sightlines where CO and OH were observed and not detected in emission (16/93); or (b) sightlines where CO was not observed but OH was observed but not detected (18/93 sightlines). In both cases we require that OH be undetected in the 1667 MHz line to a detection limit of \NOH$<$1$\times$10$^{13}$cm$^{-2}$ \citep[see][]{Li2018}, which excludes some positions with weaker continuum background sources. 
We may confidently assume that these sightlines contain very little or no \h2, and note that all but one of them lie outside the Galactic plane ($|b|$$>$10$^{\circ}$). Figure \ref{fig:src_examples} shows example maps of the immediate vicinity of 4 of these sightlines in \t353 and \ebv. Identical maps for the remaining 30 atomic sightlines, as well as all 19 sightlines with OH detections (see also Section \ref{sec:xoh}), are shown in the appendices. 

In all following subsections, \NH\ is therefore taken to be equal to \NHI, the opacity-corrected \hi\ column density, as derived along sightlines with no molecular gas detected in emission.

\subsection{ \NH\ from dust optical depth \t353}
\label{subsec:nh-from-t353}
We adopt the all-sky map of dust optical depth \t353 computed by PLC2014a. This was derived from a MBB empirical fit to IRAS and Planck maps at 3000, 857, 545 and 353 GHz, described by the expression:
\begin{equation}
I_{\nu} = \tau_{353}B_{\nu}(T_\mathrm{dust})\left(\frac{\nu}{353}\right)^{\beta_\mathrm{dust}}.
\label{eq_mbbfit}
\end{equation}

\noindent Here, \t353, dust temperature, $T_\mathrm{dust}$, and spectral index, $\beta_\mathrm{dust}$, are the three free parameters, and $B_{\nu}(T_\mathrm{dust})$ is the Planck function for dust at temperature $T_\mathrm{dust}$ which is, in this model, considered to be uniform along each sightline (see PLC2014a for more details). The relation between dust optical depth and total gas column density can then be written as:
\begin{equation}
\tau_{353} = \frac{I_{353}}{B_{353}(T_\mathrm{dust})} = \kappa_{353}r\mu m_{H}N_{H} = \sigma_{353}N_{H},
\label{eq_t353_nh}
\end{equation}

\noindent where \s353\ is the dust opacity, \k0 is the dust emissivity cross-section per unit mass (cm$^{2}$g$^{-1}$), $r$ is the dust-to-gas mass ratio, $\mu$ is the mean molecular weight and $m_{H}$ is the mass of a hydrogen atom.

\begin{figure}[h]
\centering
\includegraphics[width=1.0\linewidth]{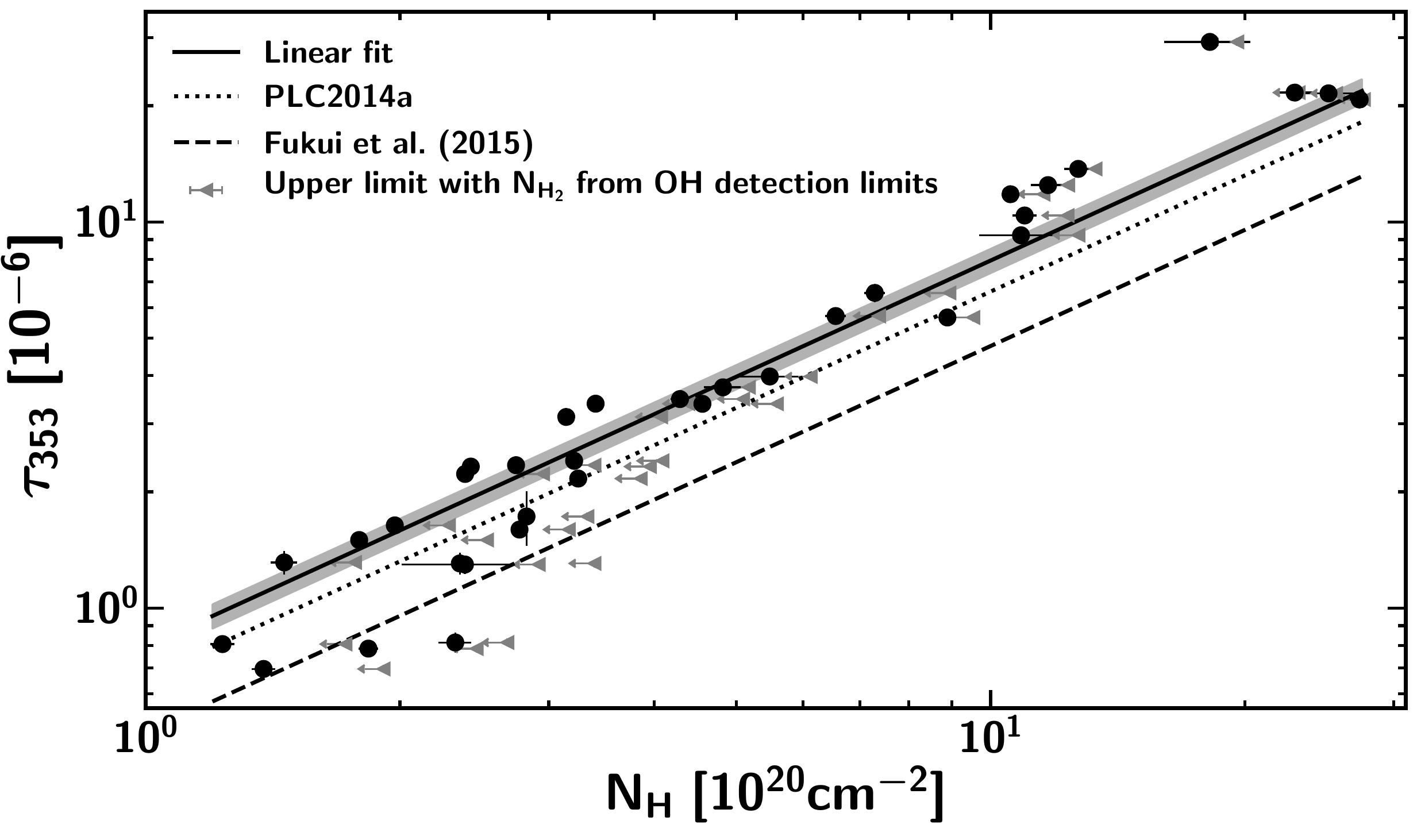}
\caption{\t353\ vs \NH\ along the 34 purely atomic sightlines described in the text. Grey triangles indicate the upper limits for \NH\ along these 34 atomic sightlines with \NHm\ estimated from the 3$\sigma$ OH detection limits using an abundance ratio \NOH/\NHm=10$^{-7}$. The thick solid line shows the linear fit to the data in this work, the dotted-line shows the conversion factor derived by PLC2014a, and the dashed line shows the conversion factor derived by \citet{Fukui2015}. (Note that all these works use the same \t353 map). \t353 error bars are from the uncertainty map of PLC2014a; the shaded region represents the 95\% confidence intervals for the linear fit, estimated from pair bootstrap resampling.}
\label{fig:tau_vs_nhi}
\end{figure}

Figure \ref{fig:tau_vs_nhi} shows the correlation between \NH\ and \t353. A tight linear trend can be seen with a Pearson coefficient of 0.95. The value of \s353\ from the orthogonal distance regression \citep{Boggs1990} linear fit is (7.9$\pm$0.6)\sigUnit\ (the intercept is set to 0), where the quoted uncertainties are the 95\% confidence limits estimated from pair bootstrap resampling. This is consistent to within the uncertainties with that obtained by PLC2014a based on all-sky \hi\ data from LAB, (6.6$\pm$1.7)\sigUnit. Note that here we have quoted the PLC2014a measurement made towards low \NHI\ positions, because the lack of any \hi\ opacity correction in that work makes this value the most reliable. However, our fit is consistent with all of \s353\ values presented in that work (which was based on the Planck R1.20 data release), to within the quoted uncertainties.

Small systematic deviations from the linear fit, evident at the high and low column density ends of the plot, are discussed further in $\S$\ref{subsec:evolution-of-dust}. 

In order to examine the possible contribution of molecular gas to \NH\ along the 34 atomic sightlines, we estimate upper limits on \NHm\ from the 3$\sigma$ OH detection limits using an abundance ratio of \xoh=$10^{-7}$ (see Section \ref{sec:xoh}). These values are tabulated in Table \ref{table:OH-parameters}, and the resulting upper limits on \NH\ are shown as gray triangles in Figure \ref{fig:tau_vs_nhi}. As expected, the \s353\ obtained from the fit to these upper limits is lower, at (6.4$\pm$0.3)\sigUnit. However, while some molecular gas may indeed be present at low levels, these limits should be considered as extreme upper bounds on the true molecular column density. 
This is particularly true for the most diffuse sightlines with the lowest column density (\NHI$<$5\nhiUnit), where the observational upper limits may appear to raise \NH\ by up to $\sim$50\%. Molecules are not expected to be well-shielded at such low columns (and indeed even CNM is largely absent along these sightlines in our data). Even for higher column density datapoints, it can be readily seen from Figures \ref{fig:src_examples} and \ref{fig:apdFig1}--\ref{fig:apdFig3} that all sightlines considered in this analysis lie well away from even the faintest outskirts of CO-bright molecular gas complexes. 
We also note that the deviations from the linear fit that will be discussed in more detail below could not be removed by any selective addition of molecular gas at levels up to these limits.

We next compare our results with the dust opacity \s353\ derived by \citet{Fukui2015} (plotted on Figure \ref{fig:tau_vs_nhi} as a dashed line). These authors derived a smaller value than in the present work (by a factor of $\sim$1.5), by restricting their fit to only the warmest dust temperatures, under the assumption that these most reliably select for genuinely optically thin \hi. They then applied this factor to the Planck \t353 map (excluding $|b|$$<$15$^{\circ}$ and CO-bright sightlines) to estimate \NHI, assuming that the contribution from CO-dark \h2 was negligible. This resulted in \NHI\ values $\sim$2--2.5 times higher than under the optically thin assumption, and motivated their hypothesis that significantly more optically thick \hi\ exists than is usually assumed. However, we find that while the \s353\ of \citet{Fukui2015} may be a good fit to some sightlines in the very low \NHI\ regime ($\lesssim$3\nhiUnit), it overestimates \NHI\ at larger column densities by $\sim$50\%. Indeed, as will be discussed below, \s353 is not expected to remain constant as dust evolves. This (combined with some contribution from CO-dark \h2) may reconcile the apparent discrepancy between their findings and absorption/emission-based measurements of the opacity-corrected \hi\ column. 

\subsection{ \NH\ from dust reddening \ebv}
\label{subsec:nh-from-ebv}
Reddening caused by the absorption and scattering of light by dust grains is defined as:
\begin{equation}
\begin{split}
E(B{-}V) = \frac{A_{V}}{R_{V}} = 1.086\frac{\kappa_{V}}{ R_{V} } r \mu  m_{\rm H} N_{\rm H}
\end{split}
\end{equation}

\noindent where \av\ is the dust extinction, $R_V$ is an empirical coefficient correlated with the average grain size and all other symbols are defined as before. In the Milky Way, $R_{V}$ is typically assumed to be 3.1 \citep{Schultz1975}, but it may vary between 2.5 and 6.0 along different sightlines \citep{Goodman1995,Draine2003}.

The ratio $\langle$\NH/\ebv$\rangle$=5.8$\times$10$^{21}$cm$^{-2}$mag$^{-1}$ \citep{Bohlin1978} is a widely-accepted standard, used in many fields of astrophysics to connect reddening measurements to gas column density. This value was derived from Lyman-${\alpha}$ and \h2\ line absorption measurements toward 100 stars \citep[see also][]{Savage1977}, and has been replicated over the years via similar methodology \citep[e.g.][]{shull1985, diplas1997, Rachford2009a}. However, a number of recent works using \hi\ 21cm data have found significantly higher values \citep[PLC2014a;][]{Liszt2014,Lenz2017}.

\begin{figure}
\includegraphics[width=1.0\linewidth]{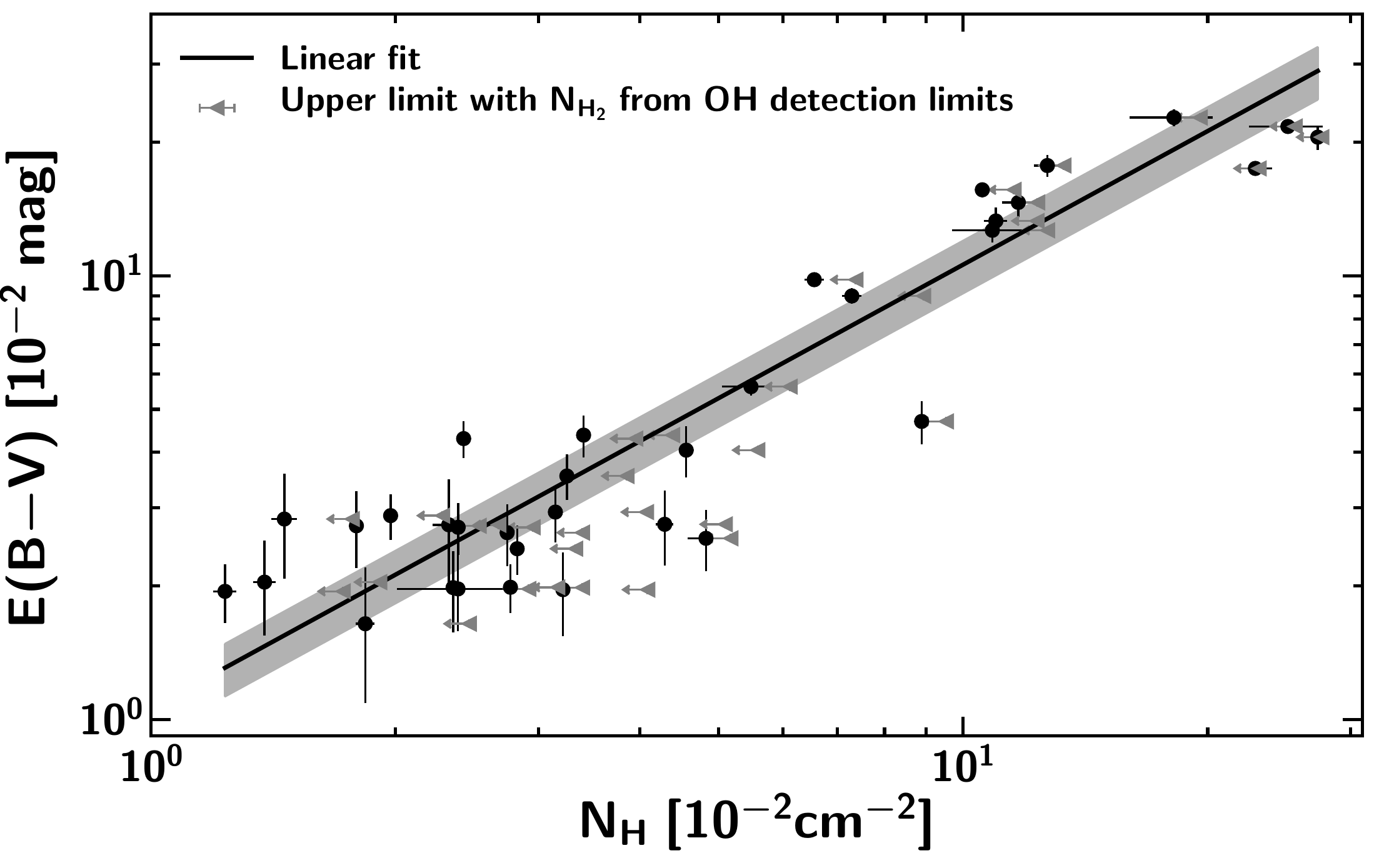}
\caption{Correlation between \NH\ and dust reddening \ebv\ from \citet{Green2018} along 34 atomic sightlines. Grey triangles indicate the upper limits for \NH\ along these 34 atomic sightlines, with \NHm\ estimated from the 3$\sigma$ OH detection limits using an abundance ratio \NOH/\NHm=10$^{-7}$. The errorbar on \ebv\ along each sightline is the standard deviation of the 20 Markov Chain realisations of \ebv\ at infinite distance; the shaded region represents the 95\% confidence intervals for the linear fit, estimated from pair bootstrap resampling.}
\label{fig:ebv_vs_nh}
\end{figure}

Here we use the all-sky map of \ebv\ from \citet{Green2018} to estimate the ratio \NH/\ebv\ for our sample of purely atomic sightlines, at $|b|$$>$5$^{\circ}$. The results are shown in Figure \ref{fig:ebv_vs_nh}. It can be seen that \ebv\ and \NH\ are strongly linearly correlated, with a Pearson coefficient of 0.93. The ratio obtained from the linear fit is \NH/\ebv=(9.4$\pm$1.6)$\times$10$^{21}$cm$^{-2}$mag$^{-1}$ (the intercept is also set to be 0), where the quoted uncertainties are the 95\% confidence limits estimated from pair bootstrap resampling. This value is a factor of 1.6 higher than \citet{Bohlin1978}.

The value obtained here is consistent with the estimate of \citet{Lenz2017}: \NH/\ebv=8.8$\times 10^{21}$\,cm$^{-2}$\,mag$^{-1}$ (no uncertainty is given in that work). These authors compared optically thin \hi\ column density from \citet{hi4pi2016} with various estimates of \ebv\ from \citet{Schlegel1998}, \citet{Peek2010}, \citet{Schlafly2014}, PLC2014a, and \citet{Meisner2015}. We note that the estimate of Lenz et al. (2017) is only valid for \NH$<$4\nhiUnit, where it seems safe to assume that the 21 cm emission is optically thin. Our value is also close to \citet{Liszt2014}, who find \NHI/\ebv=8.3$\times 10^{21}$\,cm$^{-2}$\,mag$^{-1}$ (also given without uncertainty) for $|b|$$\geq $20$^{\circ}$ and 0.015$\lesssim$\ebv$\lesssim$0.075, by comparing \hi\ data from LAB and \ebv\ from \citet{Schlegel1998}. The methodology used by these two studies differs in a number of details. For instance, \citet{Liszt2014} did not apply a gain correction to the Schlegel et al. (1998) map (whereas \citet{Lenz2017} scaled it down by 12\%), and did not smooth it to the LAB angular resolution (30$'$). However, \citet{Liszt2014} did apply an empirical correction factor to account for \hi\ opacity (albeit one whose effects on high-latitude sightlines was small). These details may account for the difference between the values obtained by these two otherwise similar studies.

We also note that, like the present work, these studies did not take into account the potential contribution of dust associated with the diffuse warm ionized gas (WIM). This would tend to produce a flattening of the \ebv\ vs \NHI\ relation at low \NHI\ and therefore increase the value of \NHI/\ebv\ artificially. Because we are able to accurately probe a large column density range (up to 3$\times$10$^{21}$\,cm$^{-2}$), we would naively expect our estimate of \NH/\ebv\ to be less affected by WIM bias than either \citet{Liszt2014} or \citet{Lenz2017} (which would tend to have a greater effect on lower column datapoints). While more work is needed to quantify the contribution of the WIM on dust emission/absorption measurements at low \ebv, we consider it unlikely to account for the difference between our work and historically lower measurements of the \NH/\ebv\ ratio.

Despite minor differences between these three studies, it is clear that they  point at a \NHI/\ebv\ value of ($\sim$8--9)$\times$10$^{21}$cm$^{-2}$mag$^{-1}$. This is 40--60\% higher than the traditional value of \citet{Bohlin1978}, which has been used by most models of interstellar dust as a reference point to set the dust-to-gas ratio \citep[e.g.][]{Draine2009,Jones2013}. We note that if \NH\ is replaced with upper limits (as discussed in \S\ref{subsec:nh-from-t353}), \NH/\ebv\ climbs yet higher, leaving this key conclusion unaffected.

\subsection{Disentangling the effects of grain evolution and dark gas on \s353}
\label{subsec:evolution-of-dust}
A number of studies have used the correlation between \t353 and \NH, particularly with regards to the search for dark gas \citep[e.g.][]{PLC2011, Fukui2014, Fukui2015, Reach2015}. It is clear that \t353 and \NH\ are in general linearly correlated only if \s353 is a constant. However, it is recognized that \s353 is sensitive to grain evolution, and significant variations in the ratio \NH/\t353 have been observed, particularly when transitioning to the high-density, molecular regime \citep[e.g.][]{PLC2014, PLC2015, Okamoto2017, Remy2017}. The origin of observed variations in \s353 may relate to a change in dust properties via \k0, and/or a variation in the dust-to-gas ratio $r$, but may also include a contribution due to the presence of dark gas, if this is unaccounted for in the estimated \NH. 

\begin{figure}
\includegraphics[width=1.0\linewidth]{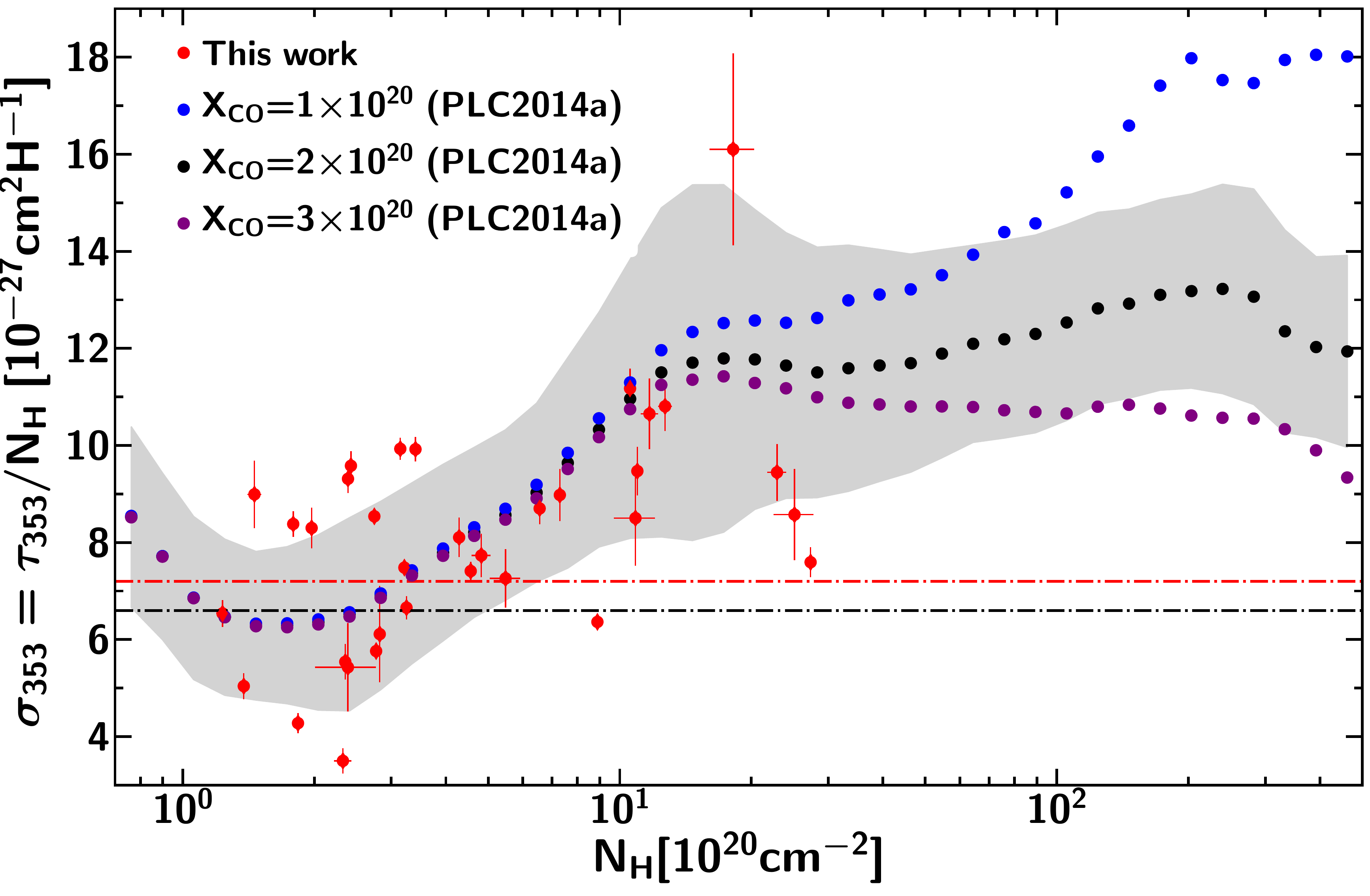}
\caption{Dust opacity \s353 versus total column density \NH\ along the 34 purely atomic sightlines presented in this work (red points), overlaid on \s353 derived for the whole sky at 30$'$ resolution from PLC2014a. Here, blue points assume an X-factor of \xco=1.0$\times$$10^{20}$, black assume \xco=2.0$\times$$10^{20}$, and violet assume \xco=3.0$\times$$10^{20}$. The grey envelope is the standard deviation of these all-sky measurements for \xco=2.0$\times$$10^{20}$. The red and black dashed lines show respectively the constant \s353\ derived from the linear fit in Section \ref{subsec:nh-from-t353} and that obtained from PLC2014a for the low column-density regime.}
\label{fig:s353_vs_nh}
\end{figure}

PLC2014a presented the variation in \s353 with \NH\ at 30$'$ resolution over the entire sky. In that work, \NH\ was derived from (\NHIthin+$X_\mathrm{CO}W_\mathrm{CO}$), thus dark gas (both optically thick \hi\ and CO-dark \h2) was unaccounted for. We reproduce their data in Figure \ref{fig:s353_vs_nh}. It can be seen that \s353 is roughly flat and at a minimum in a narrow, low-column density range \NH=(1$-$3)\nhiUnit, then increases linearly until \NH=15\nhiUnit, by which point it is almost a factor of 2 higher. It then remains approximately constant for the canonical value of \xco=2.0$\times$10$^{20}$cm$^{-2}$K$^{-1}$km$^{-1}$s. A key issue for dark gas studies is disentangling how much of the initial rise in \s353 is due to changing grain properties, and how much is due to the contribution of unseen material, whether it be opaque \hi\ or diffuse \h2. (Note also the upturn in \s353 seen at the lowest \NH, which may be due to the presence of unaccounted-for protons in the warm ionized medium.)

\begin{figure}
\includegraphics[width=1.0\linewidth]{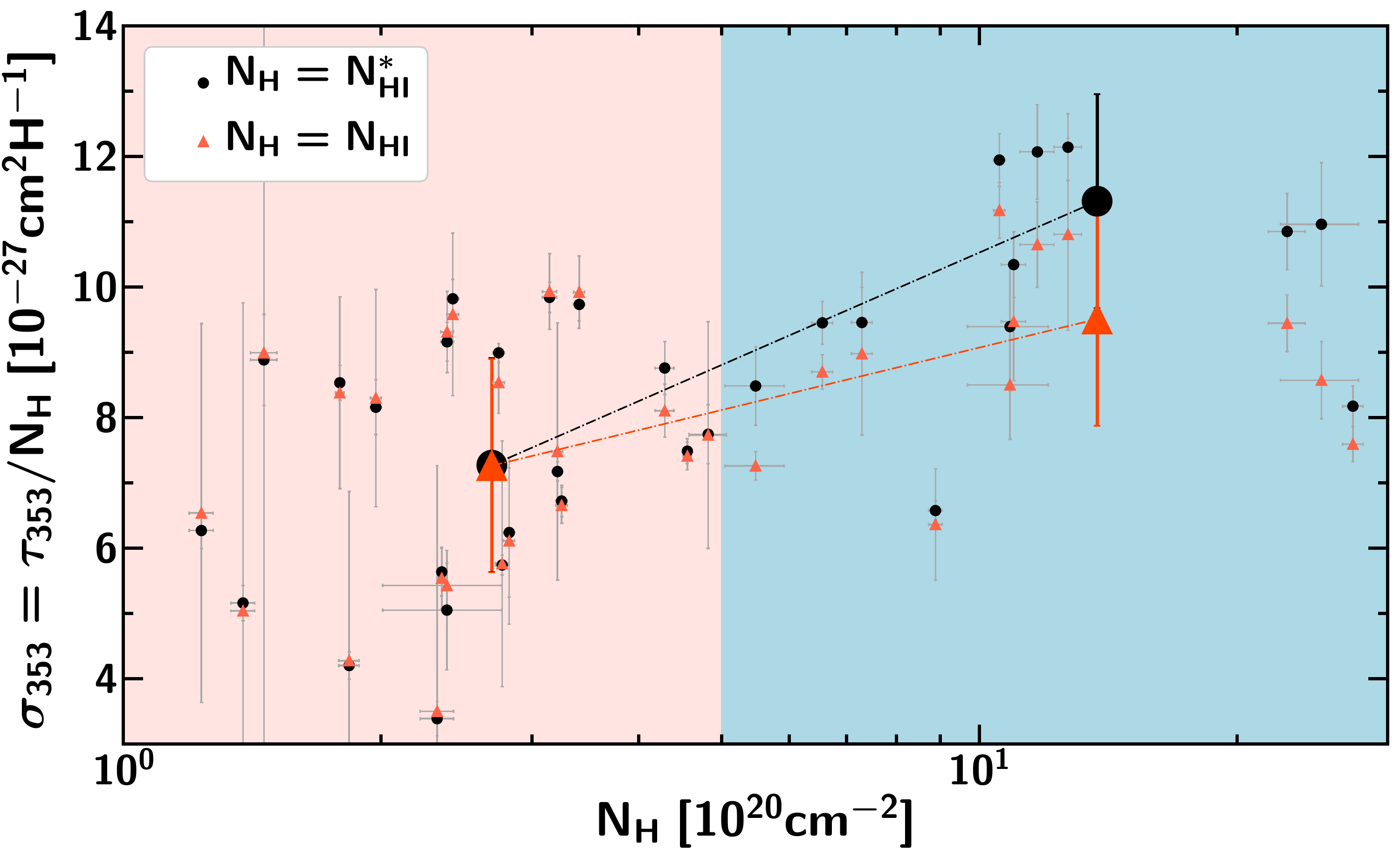}
\caption{Dust opacity \s353 versus total column density \NH\ along the 34 purely atomic sightlines presented in this work using true \NHI\ (red) and \NHIthin\ (black) as the total gas column density \NH. The vertical line is at 5\nhiUnit. The large datapoints are the average values for the low-density (\NH$<$5\nhiUnit) and high-density (\NH$>$5\nhiUnit) regions (error bars on these points are the standard error on the mean). Note that two datapoints, one black (\s353=27.2\sigUnit) and one red (\s353=16.1\sigUnit), at \NH=18.2\nhiUnit\ are not shown, but are included in the averages.}
\label{fig:s353_vs_nhi}
\end{figure}

The column density range probed by our purely atomic sightlines, \NH=(1$\sim$30)\nhiUnit\ well samples the range where \s353 undergoes its first linear increase. Dark gas is also fully accounted for in our data, since \hi\ is opacity-corrected, and no molecular gas is detected in emission along these sightlines. To quantify the effect of ignoring \hi\ opacity on \s353 we compare \s353\ deduced from the true, opacity-corrected \NHI\ with that deduced under the optically thin assumption. The results are shown in Figure \ref{fig:s353_vs_nhi}. In low column density regions (\NH$<$5\nhiUnit), each \s353\ pair from \NHI\ and \NHIthin\ are comparable. However, at higher column densities (\NH$>$5\nhiUnit) \s353\ from true \NHI\ is systematically lower than that measured from \NHIthin. On average, \s353\ obtained from optically thin \hi\ column density increases by $\sim$1.6 when going from low to high column density regions; whereas \s353\ from true \NHI\ increases by $\sim$1.4. This suggests that if \hi\ opacity is not explicitly corrected for, it can account for around one third (1/3) of the increase of \s353\ observed during the transition from diffuse to dense atomic regimes. The remaining of two thirds (2/3) must arise due to changes in dust properties.

From Equation \ref{eq_t353_nh}, we see that \s353 is a function of the dust-to-gas mass ratio, $r$, and the dust emissivity cross-section, \k0, which depends on the composition and structure of dust grains. Given the uncertainties on the efficiency of the physical processes involved in the evolution of interstellar dust grains, it is difficult at this point to conclude if the variations of \s353 observed here are due to an increase of the dust mass (i.e., $r$) or to a change in the dust emission properties (i.e., \k0). 
Using the dust model of \citet{Jones2013}, \citet{Ysard2015} suggest that most of the variations in the dust emission observed by Planck in the diffuse ISM could be explained by relatively small variations in the dust properties. That interpretation would favor a scenario in which the increase of \s353 from diffuse to denser gas is caused by the growth of thin mantles via the accretion of atoms and molecules from the gas phase. Even though this process would increase the mass of grains (and therefore increase $r$) the change of the structure of the grain surface would lead to a larger increase in \k0.
Alternatively, it is possible that this systematic variation of \t353/\NH\ could be due to residual large-scale systematic effects in the Planck data, or to the fact that the modified black-body model introduces a bias in the estimate of \t353. Neither of these explanations can be ruled out.

Figure \ref{fig:s353_vs_nh} shows \s353\ as a function of \NH\, superimposed on the results from PLC2014a. It can be seen that we observe a similar rise in \s353 in the column density range ($\sim$5$-$30)$\times$10$^{20}$cm$^{-2}$, but less extreme. In particular, most of our data points in the higher column density range (\NH$>$5\nhiUnit) are found below the PLC2014a trend, which is derived from the mean values of \s353\ over the whole sky in \NH\ bins. This is true even if we use \NHIthin\ rather than \NHI\ to derive \s353, indicating that optically thick \hi\ alone cannot shift our datapoints high enough for a perfect match. This is consistent with the fact that we are examining purely atomic sightlines, and likely happens because we are sampling comparatively low number densities ($n_H\lesssim 10$--100 \cc; a mixture of WNM and CNM), whereas the sample in PLC2014a includes molecular gas in the \NH\ bins, presumably with a higher \k0. However, in diffuse regions with \NH$<$5\nhiUnit, the mean value of \s353 from our sample is comparable with that from PLC2014a.

\subsection{\ebv\ as the more Reliable Proxy for \NH?}

We have seen that along 34 atomic sightlines \ebv\ shows a tight linear correlation with \NH\ in the column density range \NH=(1$\sim$30)\nhiUnit. \t353\ also shows a good linear relation with \NH\, but with systematic deviations as described above. 

\begin{figure}
\includegraphics[width=1.0\linewidth]{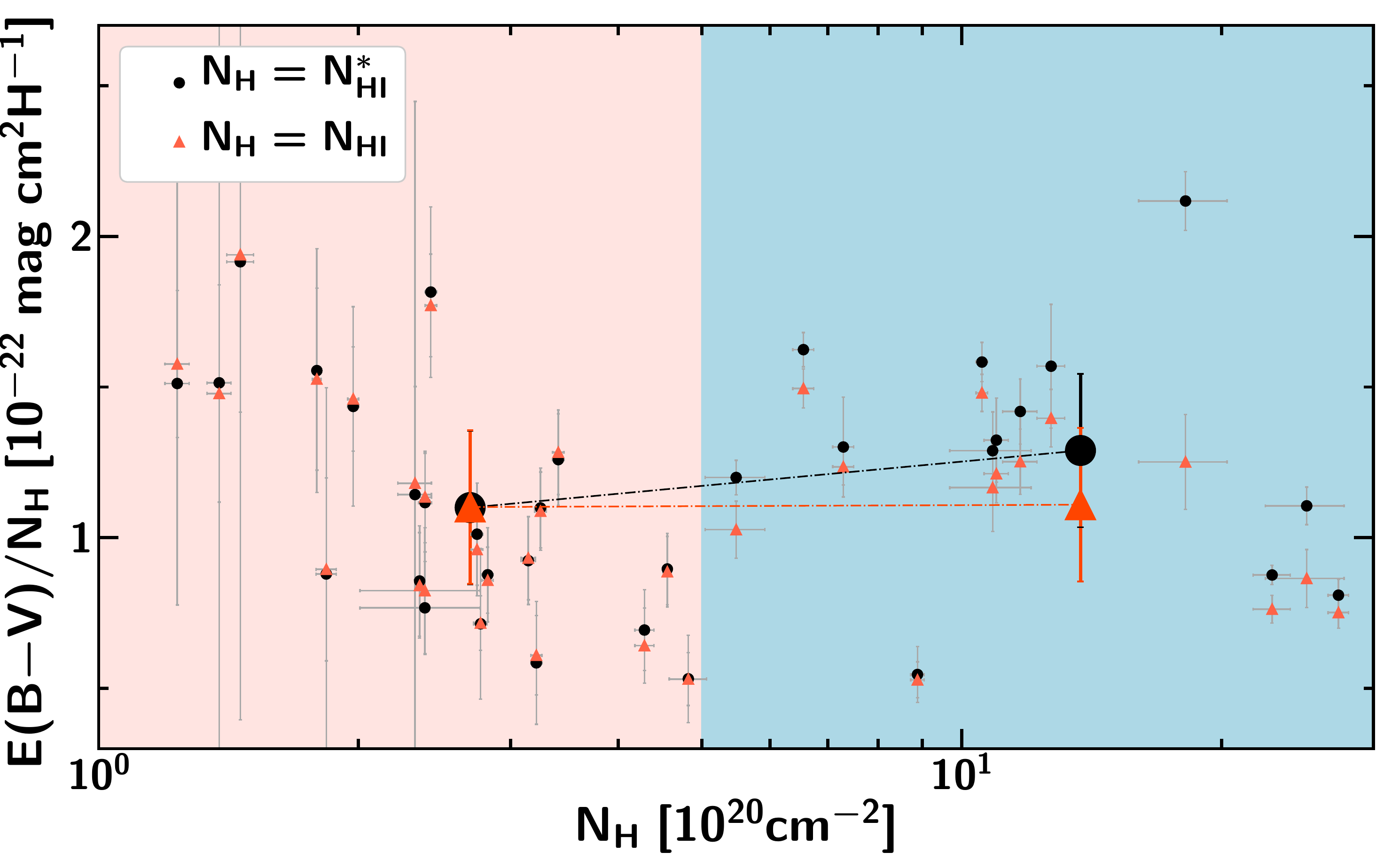}
\caption{The ratio \ebv/\NH\ as a function of \NH\ along the 34 purely atomic sightlines presented in this work, using true \NHI\ (red) and \NHIthin\ (black) as the total gas column density \NH. The dashed line is at 5\nhiUnit. The large datapoints are the average values for the low-density (\NH$<$5\nhiUnit) and high-density (\NH$>$5\nhiUnit) regions (error bars on these points are the standard error on the mean).}
\label{fig:sigA_vs_nh}
\end{figure}

Figure \ref{fig:sigA_vs_nh} replicates Figure \ref{fig:s353_vs_nhi} but for \ebv\ rather than \t353. Although the sample used here is small, these figures demonstrate clearly that the ratio \ebv/\NH\ is more stable than \t353/\NH\ over the range of column densities and sightlines covered by our analysis. In fact, with \NH\ corrected for optical depth effects, our data are compatible with a constant value for \ebv/\NH, up to \NH=30\nhiUnit. On the other hand, we have observed an increase of \t353/\NH\ with \NH\ which we suggest may be due to an increase of the dust emissivity (an increase of $r$ and/or \k0 without affecting significantly the dust absorption cross-section). While we are unfortunately unable to follow how these relations evolve at higher \av\ and in molecular gas, our results nevertheless suggest that the \ebv\ maps of \citet{Green2018} are a more reliable proxy for \NH\ than the current release of \textit{Planck} \t353 in low-to-moderate column density regimes. 

\section{OH abundance ratio \xoh}
\label{sec:xoh}
The rotational lines of CO are widely used to probe the physical properties of \h2 clouds, but in diffuse molecular regimes where CO is not detectable in emission, other species and transitions must be considered as alternative tracers of \h2. Among these, the ground-state main lines of OH are a promising dark gas tracer; they are readily detectable in translucent/diffuse molecular clouds \citep[e.g.][]{Magnani1990,Barriault2010}, and since OH is considered as a precursor molecule necessary for the formation of CO in diffuse regions \citep{Black1977,Barriault2010}, it is expected to be abundant in low-CO density/abundance regimes. 

The utility of OH as a tracer of CO-dark \h2\ depends on our ability to constrain the OH/\h2\ abundance ratio, \xoh=\NOH/\NHm. From an observational perspective, this requires good estimates of both the OH and \h2\ column densities, the latter of which often cannot be observed directly. Many efforts (both modeling and observational) have been devoted to deriving \xoh\ in different environmental conditions, which we summarize below:

\begin{itemize}
\item Astrochemical models by \citet{Black1977} found \xoh$\sim$$10^{-7}$ for the case of $\zeta$ Ophiuchi cloud.

\item 19 comprehensive models of diffuse interstellar clouds with \nh\ from 250 to 1000 \cc, $T_\mathrm{k}$ from 20 to 100K and \av\ from 0.62 to 2.12 mag \citep{vanDishoeck1986} found OH/\h2\ abundances from 1.6$\times$$10^{-8}$ to 2.9$\times$$10^{-7}$.

\item The OH abundance with respect to \h2\ from chemical models of diffuse clouds was found to vary from 7.8$\times$$10^{-9}$ to 8.3$\times$$10^{-8}$ with \av=(0.1$-$1) mag, $T_{K}$=(50$-$100) K, $n$=(50$-$1000) \cc\ \citep{Viala1986}.

\item 6 model calculations (that differ in depletion factors of heavy elements and cosmic ray ionization rate) by \citet{Nercessian1988} towards molecular gas in front of the star HD 29647 in Taurus found OH/\h2\ ratios between 5.3$\times$$10^{-8}$ and 2.5$\times$$10^{-6}$. 

\item From OH observations towards high-latitude clouds using the 43m NRAO telescope, \citet{Magnani1988} derived \xoh\ values between 4.8$\times$$10^{-7}$ to 4$\times$$10^{-6}$ in the range of \av=(0.4$-$1.1) mag assuming that \NHm=9.4$\times$$10^{20}$\av. However, we note that the excitation temperatures of the OH main lines were assumed to be equal, T$_{\mathrm{ex,1665}}$=T$_{\mathrm{ex,1667}}$, likely resulting in overestimation of \NOH\ \citep[see][]{Crutcher1979, Dawson2014}.

\item \citet{Andersson1993} obtained an OH abundance of $\sim$$10^{-7}$ from models of halos around dark molecular clouds. 

\item Combining \NOH\ data from \citet{Roueff1996} and \citet{Felenbok1996} with measurements of \NHm\ from \citet[][using UV absorption]{Savage1977}, \citet[][using UV absorption]{Rachford2002} and \citet[][using CO emission]{Joseph1986}, \citet{Liszt2002} find \xoh=(1.0$\pm$0.2)$\times$$10^{-7}$ toward diffuse clouds.

\item \citet{Weselak2010} derived OH abundances of (1.05$\pm$0.24)$\times$$10^{-7}$ from absorption line observations of 5 translucent sightlines, with molecular hydrogen column densities \NHm\ measured through UV absorption by \citep{Rachford2002, Rachford2009a}.

\item \citet{Xu2016} report that \xoh\ decreases from 8$\times$10$^{-7}$ to 1$\times$10$^{-7}$ across a boundary region of the Taurus molecular cloud, over the range Av=0.4$-$2.7 mag. \NHm\ was obtained from an integration of \av-based estimates of the \h2\ volume density (assuming \NHm=9.4$\times$$10^{20}$\av).

\item Recently, \citet{Rugel2018} report a median \xoh$\sim$1.3$\times$$10^{-7}$ from THOR Survey observations of OH absorption in the first Milky Way quadrant, with \NHm\ estimated from $^{13}$CO(1-0).
\end{itemize}

\noindent Overall, while model calculations tend to produce some variation in the OH abundance ratio over different parts of parameter space (8$\times$10$^{-9}$ to 4$\times$10$^{-6}$), observationally-determined measurements of \xoh\ cluster fairly tightly around $10^{-7}$, with some suggestion that this may decrease for denser sightlines.

In this paper we determine our own OH abundances, using the MS dataset to provide \NOH\ and \NHI; then employing \t353 and \ebv\ (along with our own conversion factors) to compute molecular hydrogen column densities as $N_{\mathrm{H}_{2}}$=$\frac{1}{2}(N_\mathrm{H}{-}N_\mathrm{HI})$. We note that since this dust-based estimate of \NHm\ cannot be decomposed in velocity space, the OH abundances are determined in an integrated fashion for each sightline, and not on a component-by-component basis. While CO was detected along all but one sightline, it was not detected towards all velocity components, meaning that our abundances are generally computed for a mixture of CO-dark and CO-bright \h2 \citep[for further details see][]{Li2018}.

The OH column densities derived in Section \ref{sec:OH_data_analysis} are derived from direct measurements of \Tex\ and $\tau$. This means that they should be accurate compared to methods that rely on assumptions about these variables \citep[see e.g.][]{Crutcher1979, Dawson2014}. In computing \NH\, we assume that the linear correlations (deduced from \t353, \ebv\ and \NHI\ towards 34 atomic sightlines) still hold in molecular regions. In this manner, estimates of the OH/\h2\ abundance ratio can be obtained within a range of visual extinction \av=(0.25$-$4.8) mag. We note that of our 19 OH-bright sightlines, 5 produce \NHm\ that is either negative or consistent with zero to within the measurement uncertainties; these are excluded from the analysis.

\begin{figure}
\includegraphics[width=1.0\linewidth]{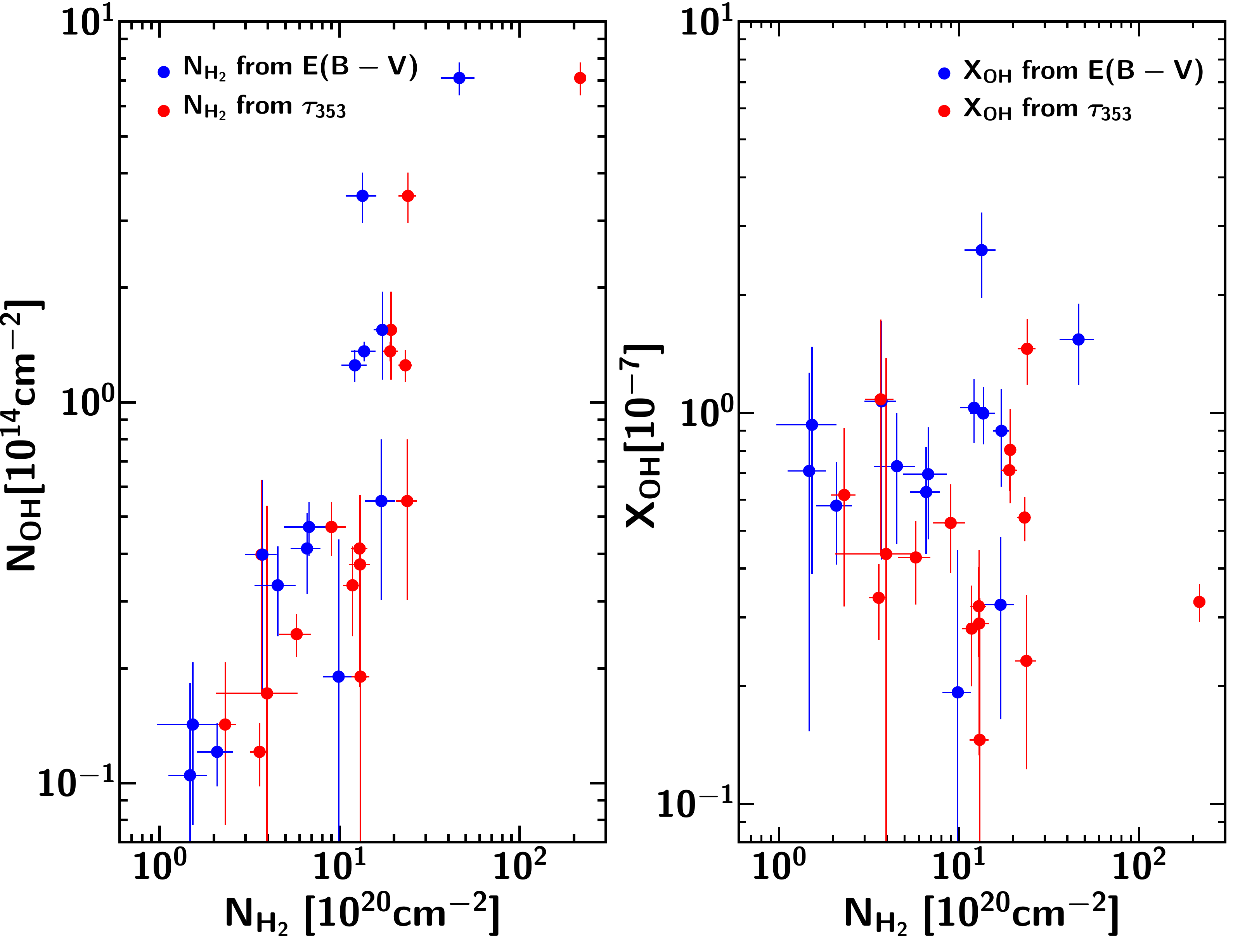}
\caption{Left: \NOH\ as a function of \NHm obtained from the two \NH\ proxies, \ebv\ (blue) and \t353 (red). Right: \xoh\ derived from the two proxies as a function of \NHm.}
\label{fig:oh_vs_h2}
\end{figure}

Figure \ref{fig:oh_vs_h2} shows \NOH\ and \xoh\ as functions of \NHm. We find \NOH\ increases approximately linearly with \NHm, and the OH/\h2\ abundance ratio is approximately consistent for the two methods, with no evidence of and systematic trends with increasing column density. Differences arise due to the overestimation of \NH\ derived from \t353 along dense sightlines compared to \NH\ from \ebv. As discussed in Section \ref{sec:proxies-for-nh}, \s353 varies by up to a factor of 2 in the range \NH=(1$\sim$30)$\times$10$^{20}$cm$^{-2}$, whereas the ratio $\langle$\NH/\ebv$\rangle$
is quite constant. The mean and standard deviation of the \xoh\ distribution deduced from \ebv\ is (0.9$\pm$0.6)$\times$10$^{-7}$, which is close to the canonical value of $\sim$1$\times$10$^{-7}$, and double the \xoh\ from \t353, (0.5$\pm$0.3)$\times$10$^{-7}$. We regard the higher value as more reliable. 

\section{Conclusions}
\label{sec:conclusion}

We have combined accurate, opacity-corrected \hi\ column densities from the Arecibo Millennium Survey and 21-SPONGE 
with thermal dust data from the \textit{Planck} satellite and the new \ebv\ maps of \citet{Green2018}. We have also made use of newly published Millennium Survey OH data and information on CO detections from \citet{Li2018}. In combination, these datasets allow us to select reliable subsamples of purely-atomic (or partially-molecular) sightlines, and hence assess the impact of \hi\ opacity on the scaling relations commonly used to convert dust data to total proton column density \NH. They also allow us to make new measurements of the OH/H$_2$ abundance ratio, which is essential in interpreting the next generation of OH datasets. Our key conclusions are as follows:

1. \hi\ opacity effects become important above \NHI$>$5\nhiUnit; below this value the optically thin assumption may usually be considered reliable.

2. Along purely atomic sightlines with \NH=\NHI=(1--30)\nhiUnit, the dust opacity, {\s353}=\t353/\NH, is $\sim$40\% higher for moderate-to-high column densities than low (defined as above and below \NH=5\nhiUnit). We have argued that this rise is likely due to the evolution of dust grains in the atomic ISM, although large-scale systematics in the \textit{Planck} data cannot be definitively ruled out. Failure to account for \hi\ opacity can cause an additional apparent rise of the order of $\sim$20\%.

3. For purely atomic sightlines, we measure a \NH/\ebv\ ratio of  (9.4$\pm$1.6)$\times$10$^{21}$cm$^{-2}$mag$^{-1}$. This is consistent with \citet{Lenz2017} and \citet{Liszt2014}, but 60\% higher than the canonical value from \citet{Bohlin1978}. 

4. Our results suggest that \NH\ derived from the \ebv\ map of \citet{Green2018} is more reliable than that obtained from the \t353 map of PLC2014a in low-to-moderate column density regimes.  

5. We measure the OH/H$_2$ abundance ratio, \xoh, along a sample of 16 molecular sightlines. We find \xoh$\sim$1$\times10^{-7}$, with no evidence of a systematic trend with column density. Since our sightlines include both CO-dark and CO-bright molecular gas components, this suggests that OH may be used as a reliable proxy for \h2\ over a broad range of molecular regimes.

\section*{Acknowledgments}
JRD is the recipient of an Australian Research Council (ARC) DECRA Fellowship (project number DE170101086). N.M.-G. acknowledges the support of the ARC through Future Fellowship FT150100024. LB acknowledges the support from CONICYT grant PFB06. We are indebted to Professor Mark Wardle for providing us with valuable advice and support. We gratefully acknowledge discussions with Dr. Cormac Purcell and Anita Petzler. Finally, we thank the anonymous referee for comments and criticisms which allowed us to improve the paper.

This research has made use of the NASA/IPAC Infrared Science Archive, which is operated by the Jet Propulsion Laboratory, California Institute of Technology, under contract with the National Aeronautics and Space Administration.

\bibliographystyle{apj}
\bibliography{dust-gas-scaling-relations}
\clearpage

\cleardoublepage

\appendix
\section{Appendix A: Locations of atomic sightlines}
\label{apdxa:atomic_sightlines}
\begin{figure*}[!htbp]
\centering
\includegraphics[width=\textwidth]{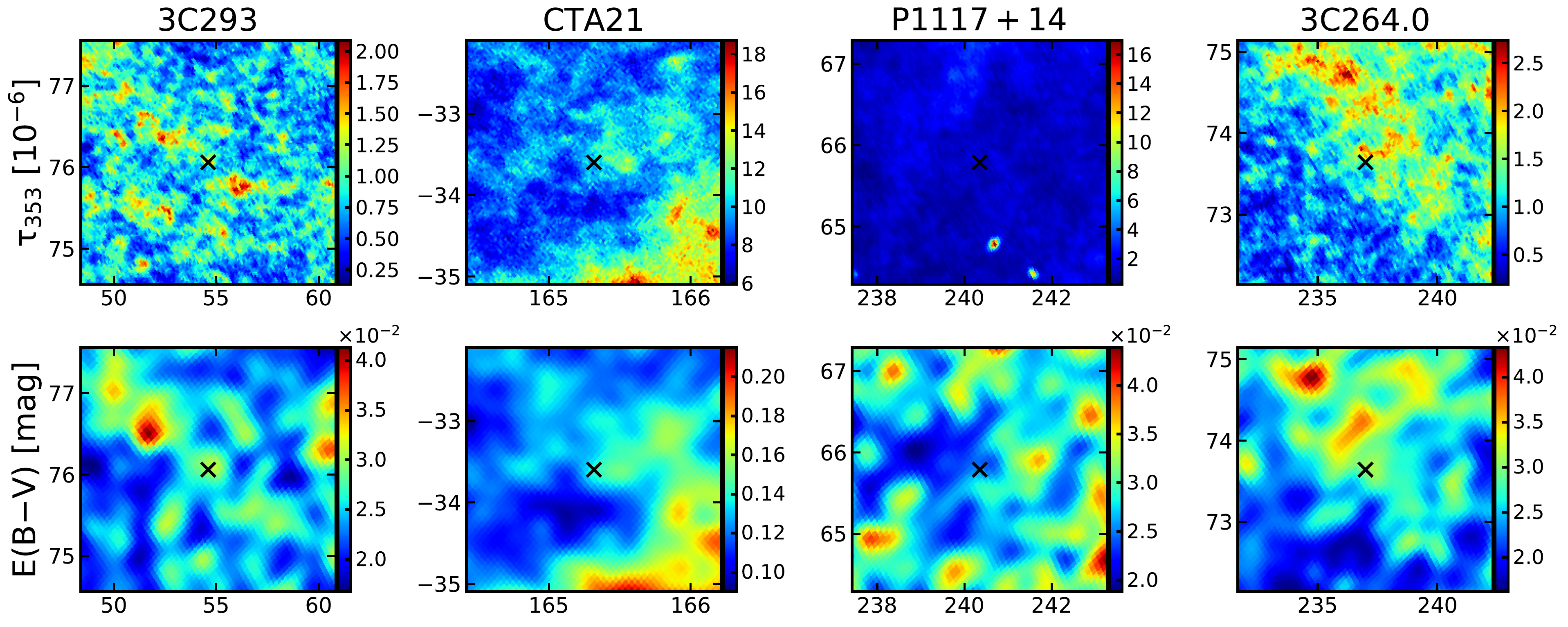}
\includegraphics[width=\textwidth]{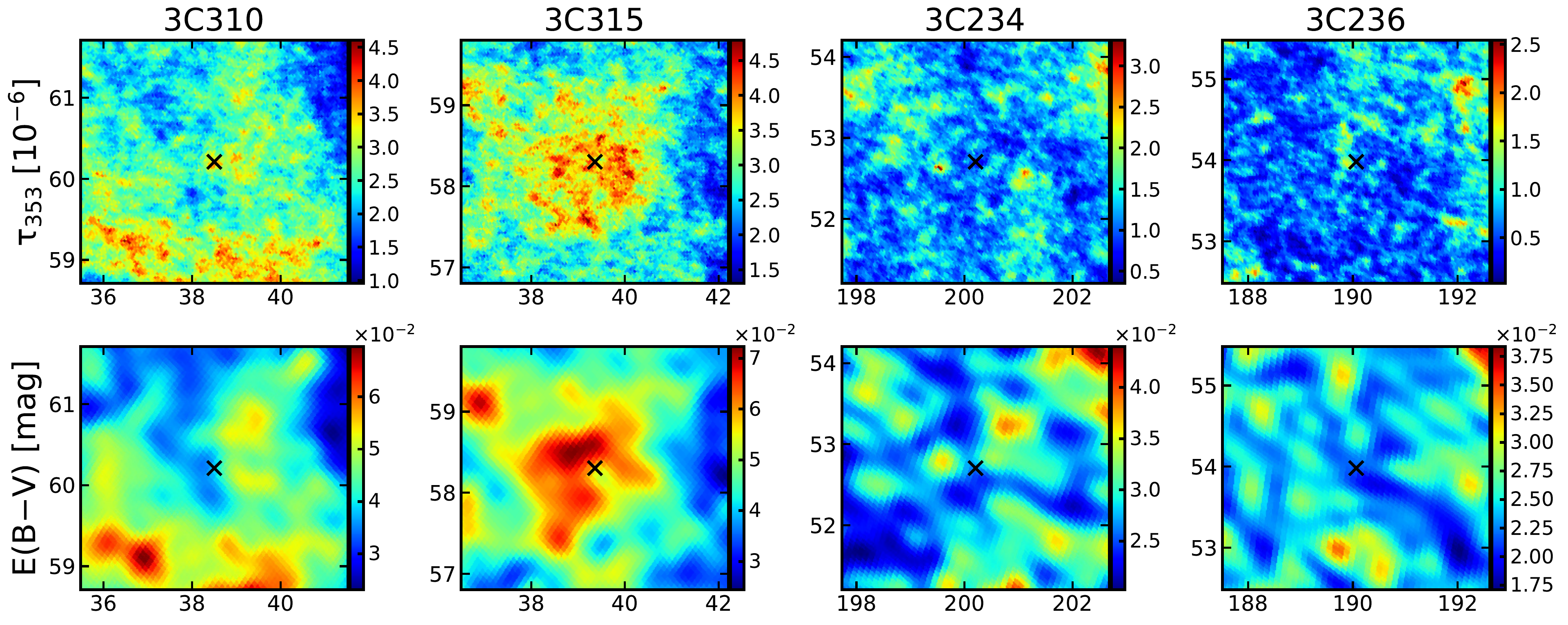}
\caption{See Figure \ref{fig:src_examples} for details.}
\label{fig:apdFig1}
\end{figure*}

\begin{figure*}[!htbp]
\centering\includegraphics[width=\textwidth]{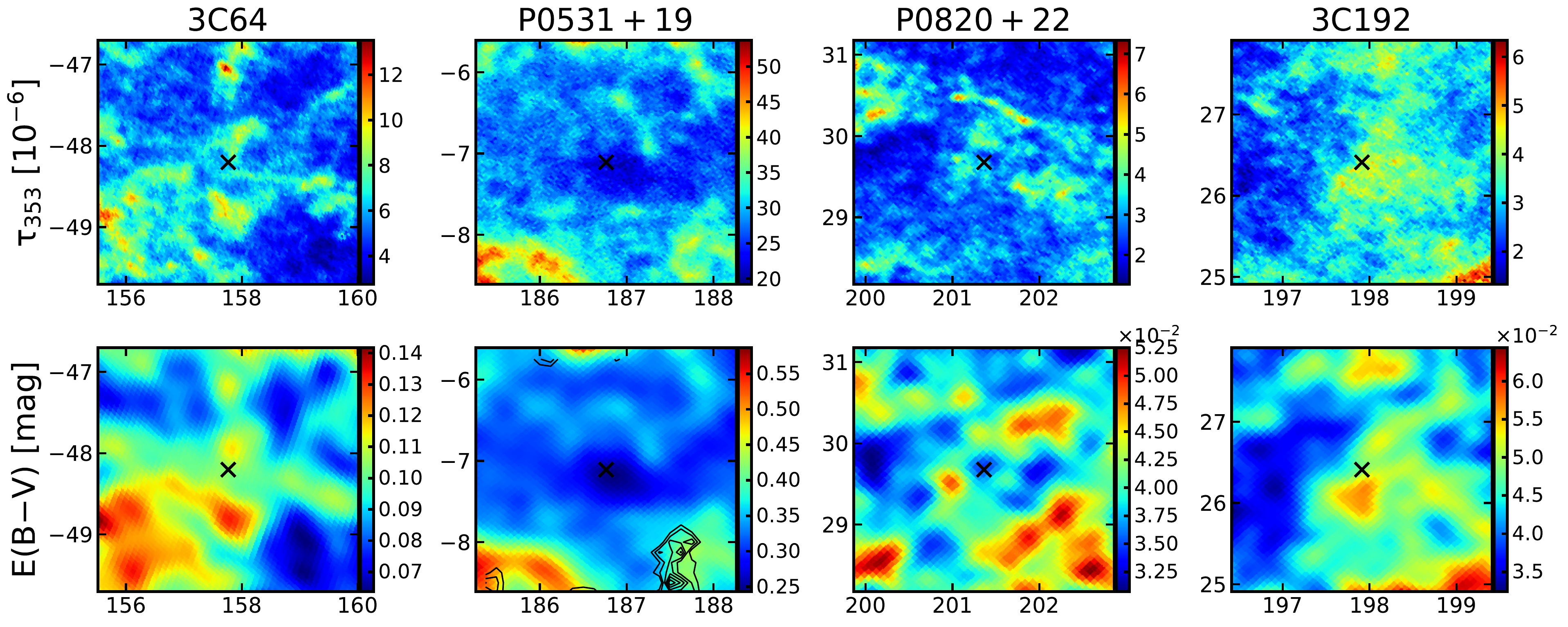}
\includegraphics[width=\textwidth]{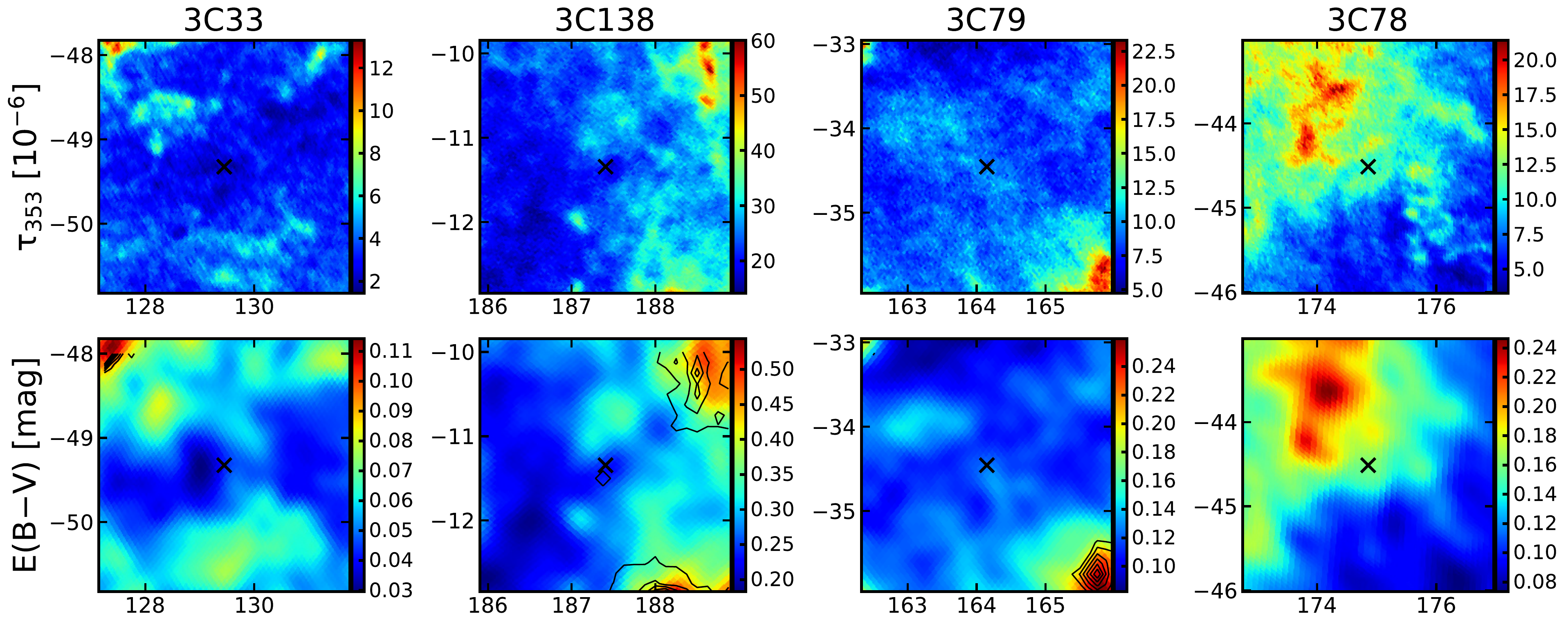}
\includegraphics[width=\textwidth]{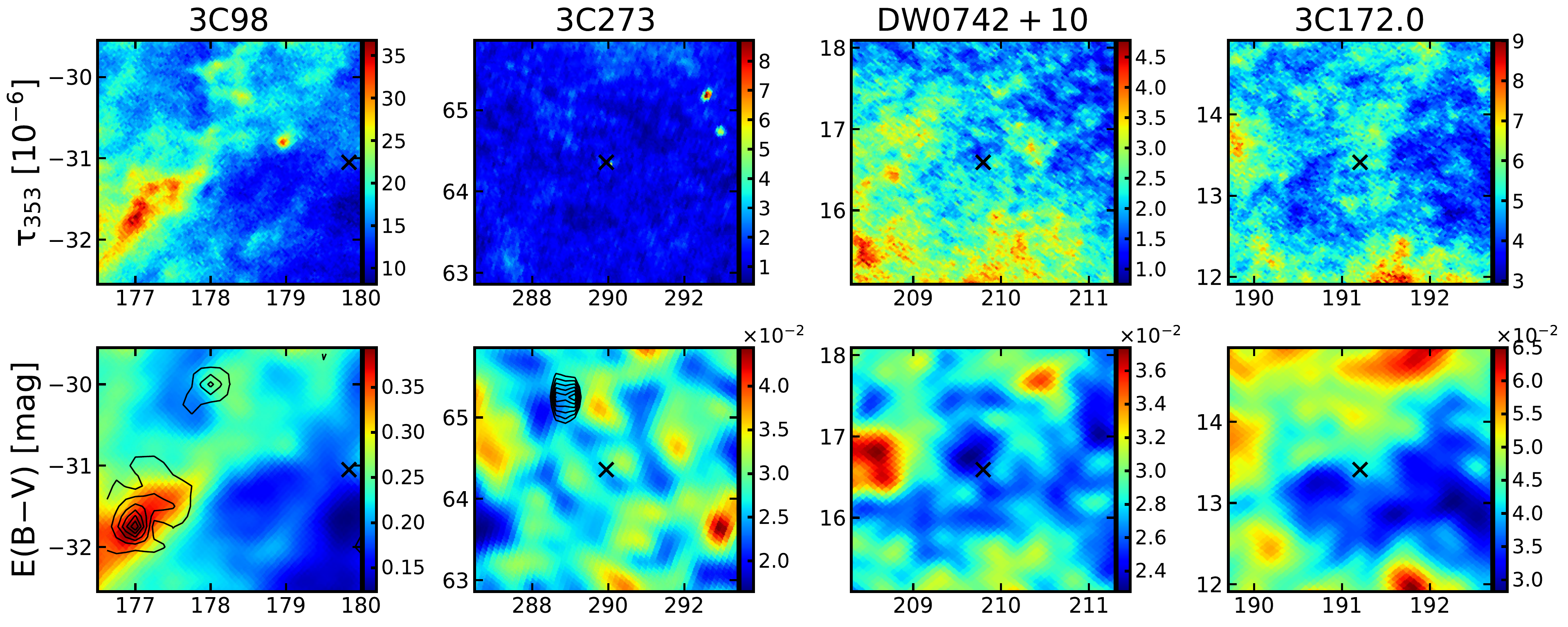}
\caption{See Figure \ref{fig:src_examples} for details.}
\label{fig:apdFig2}
\end{figure*}

\begin{figure*}[!htbp]
\centering
\includegraphics[width=0.98\textwidth]{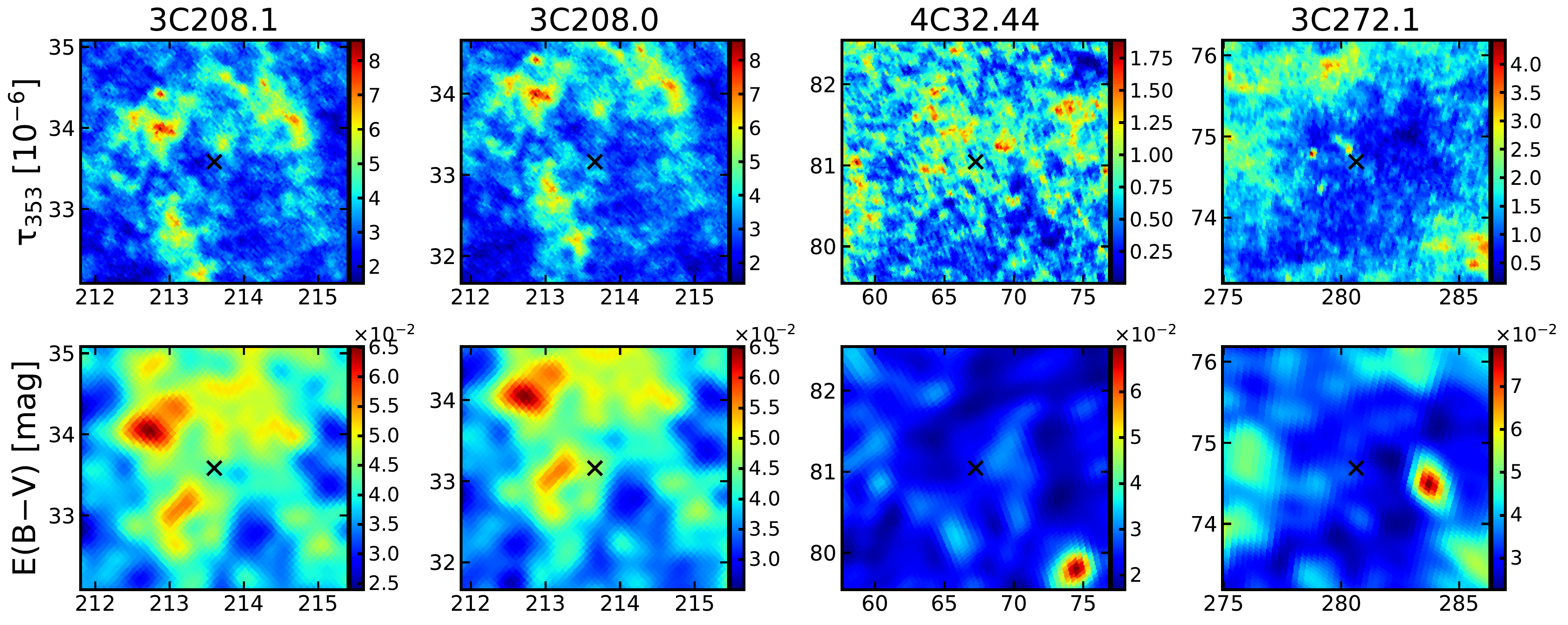}
\includegraphics[width=0.98\textwidth]{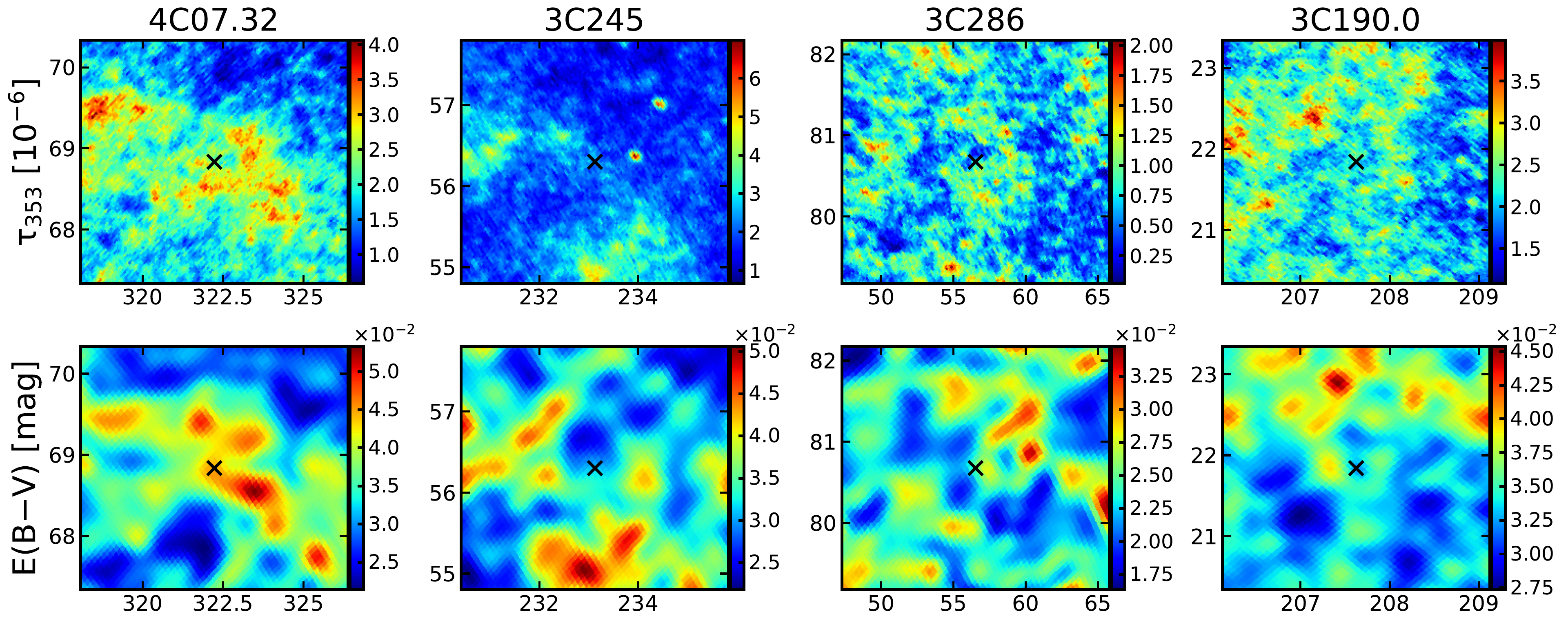}
\includegraphics[width=0.50\textwidth]{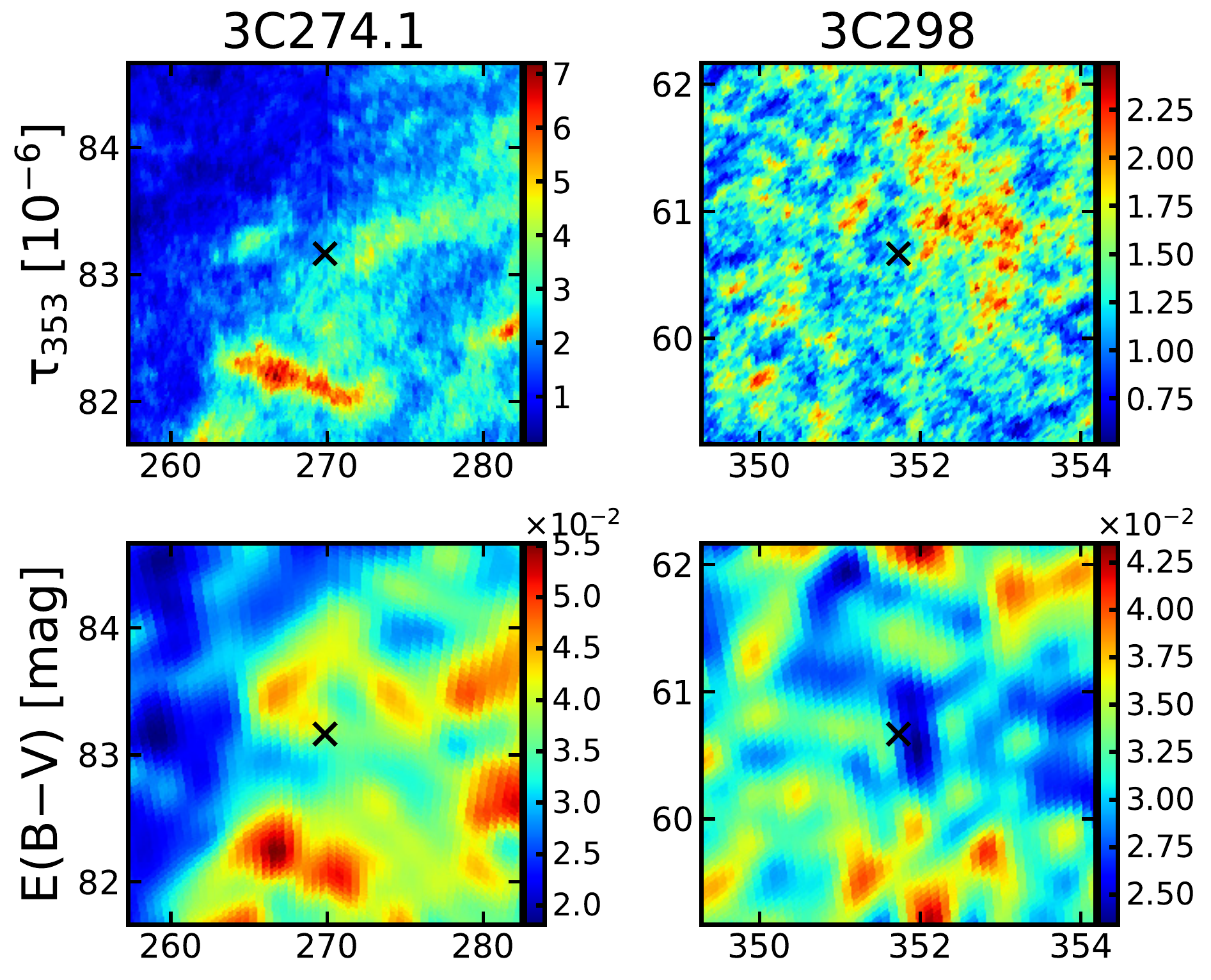}
\caption{See Figure \ref{fig:src_examples} for details.}
\label{fig:apdFig3}
\end{figure*}
\clearpage

\section{Appendix B: Locations of OH sightlines}
\label{apdx:OH_sightlines}
\begin{figure*}[!htbp]
\centering     
\includegraphics[width=\textwidth]{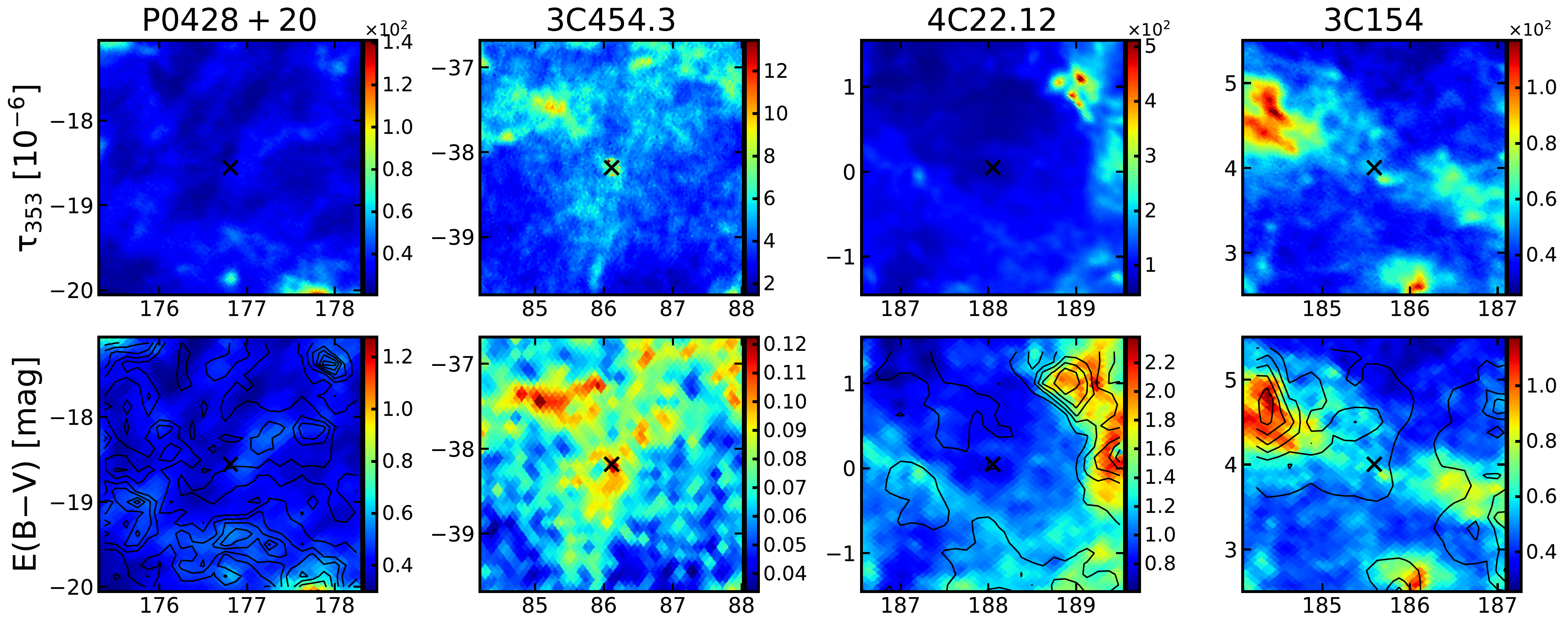}
\includegraphics[width=\textwidth]{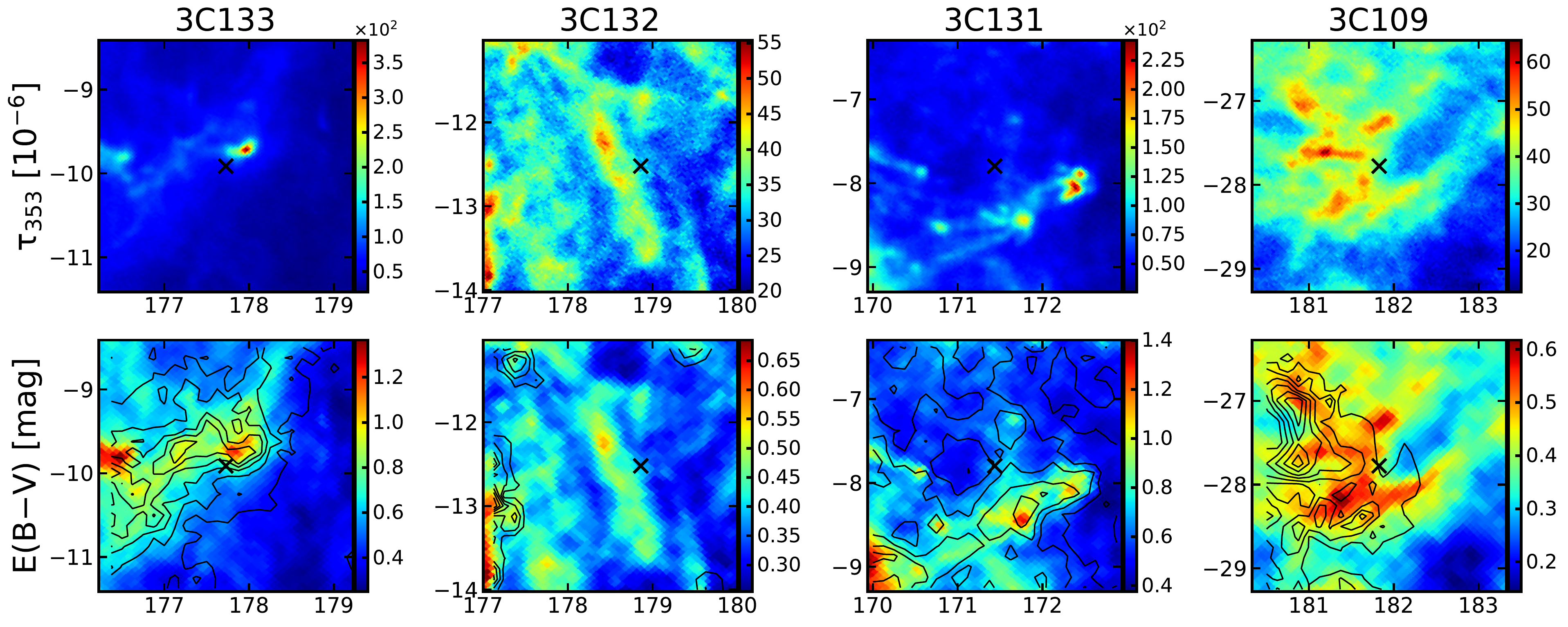}
\includegraphics[width=\textwidth]{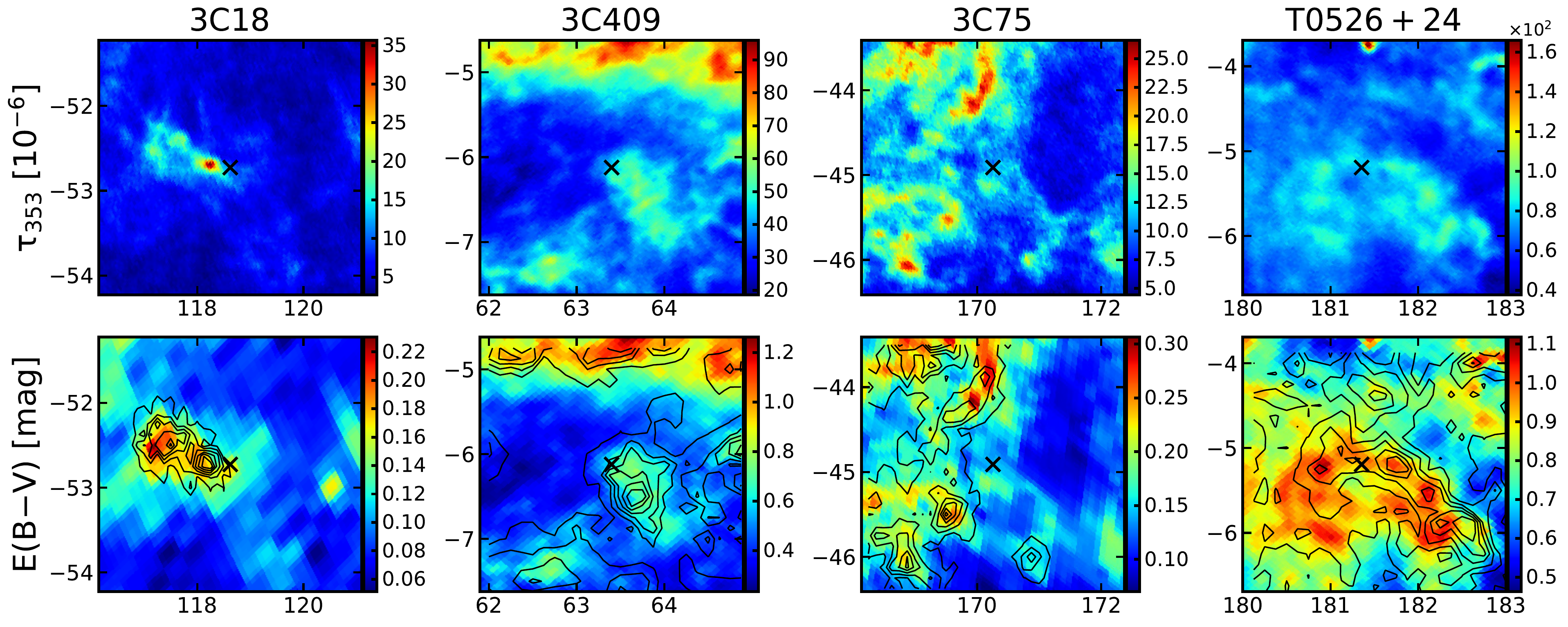}
\caption{See Figure \ref{fig:src_examples} for details.}
\label{fig:apdFig4}
\end{figure*}

\begin{figure*}[!htbp]
\centering     
\includegraphics[width=0.95\textwidth]{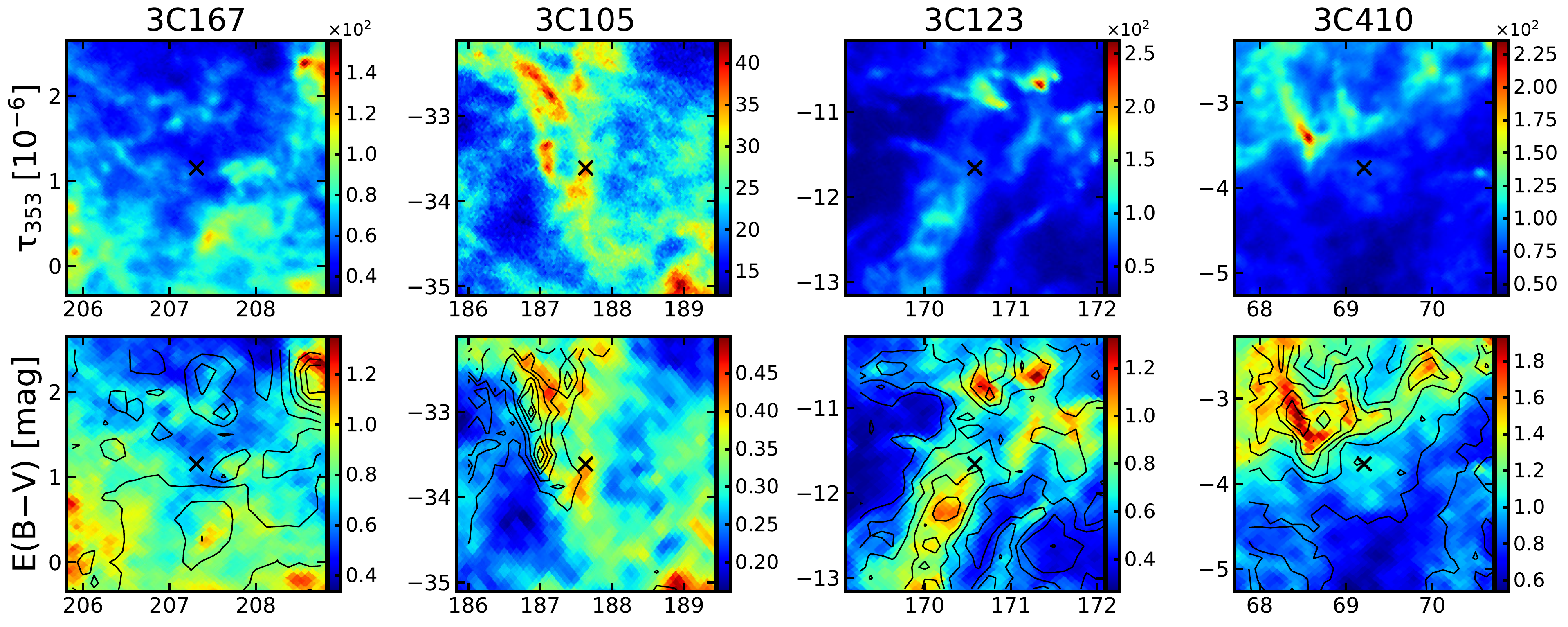}
\includegraphics[width=0.7\textwidth]{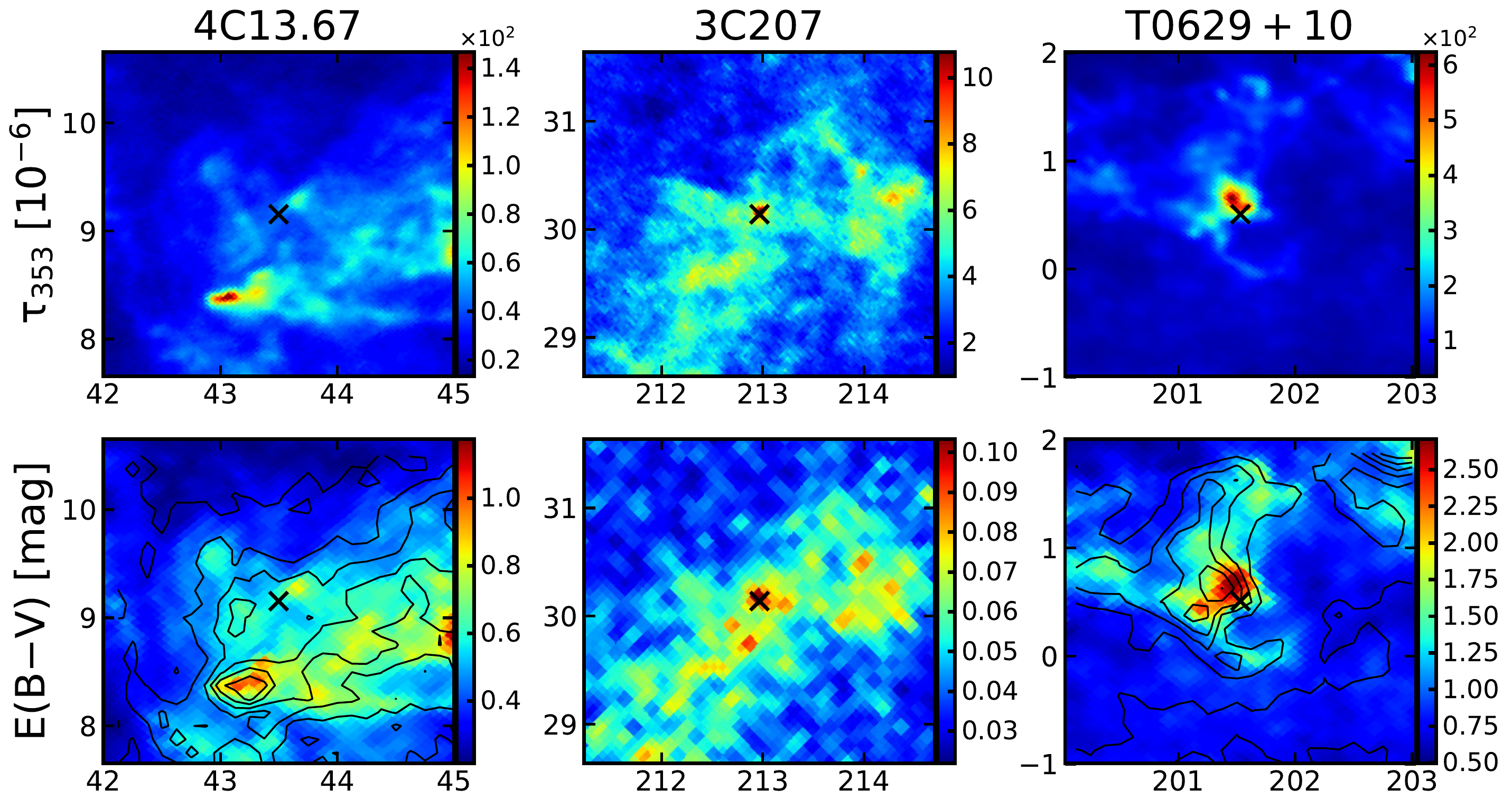}
\caption{See Figure \ref{fig:src_examples} for details.}
\label{fig:apdFig5}
\end{figure*}

\begin{sidewaystable}[h!]
\fontsize{7}{6}\selectfont
\centering
\caption{Parameters for OH main lines}
\label{table:OH-parameters}
\renewcommand{\arraystretch}{1.5}
\begin{tabular}{ lllllllclllll  }
 \hline
 \hline
\multirow{2}{*}{$Source$} & \multicolumn{1}{c}{\multirow{2}{*}{$l/b$}} & \multicolumn{6}{c}{OH(1665)} & \multicolumn{5}{c}{OH(1667)}\\
\cline{3-7}
\cline{9-13}
 & & \multicolumn{1}{c}{$\tau$} & \multicolumn{1}{c}{$V_{lsr}$} & \multicolumn{1}{c}{$\Delta V$} & \multicolumn{1}{c}{$T_{ex}$} & \multicolumn{1}{c}{$N(OH)$} & & \multicolumn{1}{c}{$\tau$} & \multicolumn{1}{c}{$V_{lsr}$} & \multicolumn{1}{c}{$\Delta V$} & \multicolumn{1}{c}{$T_{ex}$} & \multicolumn{1}{c}{$N(OH)$} \\
 
(Name) & \multicolumn{1}{c}{$(^{o})$} & \multicolumn{1}{c}{$\ $} & \multicolumn{1}{c}{$(km s^{-1})$} & \multicolumn{1}{c}{$(km s^{-1})$} & \multicolumn{1}{c}{$(K)$} & \multicolumn{1}{c}{$(10^{14}cm^{-2})$} & & \multicolumn{1}{c}{$\ $} & \multicolumn{1}{c}{$(km s^{-1})$} & \multicolumn{1}{c}{$(km s^{-1})$} & \multicolumn{1}{c}{$(K)$} & \multicolumn{1}{c}{$(10^{14}cm^{-2})$} \\ 
\hline

3C105 & 187.6/-33.6 & 0.0156 $\pm$ 0.0003 & 8.14 $\pm$ 0.01 & 0.95 $\pm$ 0.03 & 4.65 $\pm$ 1.86 & 0.29 $\pm$ 0.12 & &  0.0265 $\pm$ 0.0004 & 8.17 $\pm$ 0.01 & 0.94 $\pm$ 0.02 & 3.95 $\pm$ 0.95 & 0.23 $\pm$ 0.06 \\
3C105 & 187.6/-33.6 & 0.0062 $\pm$ 0.0003 & 10.22 $\pm$ 0.02 & 0.93 $\pm$ 0.06 & 8.5 $\pm$ 4.89 & 0.21 $\pm$ 0.12 & &  0.0104 $\pm$ 0.0004 & 10.25 $\pm$ 0.02 & 0.96 $\pm$ 0.04 & 7.66 $\pm$ 3.41 & 0.18 $\pm$ 0.08 \\
3C109 & 181.8/-27.8 & 0.0023 $\pm$ 0.0003 & 9.15 $\pm$ 0.11 & 1.03 $\pm$ 0.26 & 18.28 $\pm$ 27.06 & 0.18 $\pm$ 0.27 & &  0.0036 $\pm$ 0.0004 & 9.24 $\pm$ 0.05 & 0.75 $\pm$ 0.12 & 24.58 $\pm$ 8.7 & 0.16 $\pm$ 0.06 \\
3C109 & 181.8/-27.8 & 0.0036 $\pm$ 0.0003 & 10.45 $\pm$ 0.07 & 0.98 $\pm$ 0.15 & 13.97 $\pm$ 5.41 & 0.21 $\pm$ 0.09 & &  0.0053 $\pm$ 0.0004 & 10.55 $\pm$ 0.04 & 1.02 $\pm$ 0.1 & 13.63 $\pm$ 4.48 & 0.18 $\pm$ 0.06 \\
3C123 & 170.6/-11.7 & 0.0191 $\pm$ 0.0007 & 3.65 $\pm$ 0.06 & 1.19 $\pm$ 0.11 & 10.92 $\pm$ 3.26 & 1.05 $\pm$ 0.33 & &  0.0348 $\pm$ 0.0009 & 3.71 $\pm$ 0.04 & 1.22 $\pm$ 0.07 & 10.92 $\pm$ 2.69 & 1.1 $\pm$ 0.28 \\
3C123 & 170.6/-11.7 & 0.0431 $\pm$ 0.0023 & 4.43 $\pm$ 0.01 & 0.53 $\pm$ 0.03 & 8.06 $\pm$ 0.78 & 0.78 $\pm$ 0.09 & &  0.0919 $\pm$ 0.0029 & 4.46 $\pm$ 0.0 & 0.53 $\pm$ 0.01 & 7.7 $\pm$ 0.65 & 0.89 $\pm$ 0.08 \\
3C123 & 170.6/-11.7 & 0.0337 $\pm$ 0.0008 & 5.37 $\pm$ 0.01 & 0.91 $\pm$ 0.03 & 11.59 $\pm$ 4.3 & 1.53 $\pm$ 0.57 & &  0.0784 $\pm$ 0.0009 & 5.47 $\pm$ 0.01 & 0.92 $\pm$ 0.01 & 8.79 $\pm$ 2.57 & 1.5 $\pm$ 0.44 \\
3C131 & 171.4/-7.8 & 0.0065 $\pm$ 0.0005 & 4.55 $\pm$ 0.02 & 0.56 $\pm$ 0.05 & 12.52 $\pm$ 3.59 & 0.19 $\pm$ 0.06 & &  0.0117 $\pm$ 0.0004 & 4.64 $\pm$ 0.01 & 0.78 $\pm$ 0.04 & 6.96 $\pm$ 1.98 & 0.15 $\pm$ 0.04 \\
3C131 & 171.4/-7.8 & 0.0073 $\pm$ 0.0006 & 6.81 $\pm$ 0.06 & 2.91 $\pm$ 0.23 & 8.94 $\pm$ 4.54 & 0.82 $\pm$ 0.42 & &  0.0089 $\pm$ 0.0005 & 5.84 $\pm$ 0.03 & 0.67 $\pm$ 0.08 & 11.04 $\pm$ 2.07 & 0.16 $\pm$ 0.04 \\
3C131 & 171.4/-7.8 & 0.0166 $\pm$ 0.0007 & 6.59 $\pm$ 0.01 & 0.42 $\pm$ 0.02 & 5.69 $\pm$ 0.85 & 0.17 $\pm$ 0.03 & &  0.0319 $\pm$ 0.0007 & 6.55 $\pm$ 0.01 & 0.45 $\pm$ 0.02 & 5.99 $\pm$ 0.84 & 0.2 $\pm$ 0.03 \\
3C131 & 171.4/-7.8 & 0.0521 $\pm$ 0.0007 & 7.23 $\pm$ 0.0 & 0.55 $\pm$ 0.01 & 5.91 $\pm$ 0.64 & 0.72 $\pm$ 0.08 & &  0.0927 $\pm$ 0.0005 & 7.22 $\pm$ 0.0 & 0.65 $\pm$ 0.01 & 5.98 $\pm$ 0.34 & 0.85 $\pm$ 0.05 \\
3C132 & 178.9/-12.5 & 0.0033 $\pm$ 0.0003 & 7.82 $\pm$ 0.04 & 0.9 $\pm$ 0.1 & 15.55 $\pm$ 6.23 & 0.19 $\pm$ 0.08 & &  0.0056 $\pm$ 0.0003 & 7.79 $\pm$ 0.02 & 0.79 $\pm$ 0.06 & 23.56 $\pm$ 2.17 & 0.25 $\pm$ 0.03 \\
3C133 & 177.7/-9.9 & 0.1008 $\pm$ 0.001 & 7.66 $\pm$ 0.0 & 0.53 $\pm$ 0.0 & 4.47 $\pm$ 0.44 & 1.01 $\pm$ 0.1 & &  0.2132 $\pm$ 0.0014 & 7.68 $\pm$ 0.0 & 0.52 $\pm$ 0.0 & 3.25 $\pm$ 0.27 & 0.85 $\pm$ 0.07 \\
3C133 & 177.7/-9.9 & 0.0149 $\pm$ 0.001 & 7.94 $\pm$ 0.02 & 1.22 $\pm$ 0.04 & 7.08 $\pm$ 3.08 & 0.55 $\pm$ 0.24 & &  0.0333 $\pm$ 0.0013 & 7.96 $\pm$ 0.01 & 1.23 $\pm$ 0.02 & 4.17 $\pm$ 0.99 & 0.4 $\pm$ 0.1 \\
3C154 & 185.6/4.0 & 0.0266 $\pm$ 0.0006 & -2.32 $\pm$ 0.02 & 0.74 $\pm$ 0.03 & 2.69 $\pm$ 1.93 & 0.23 $\pm$ 0.16 & &  0.0429 $\pm$ 0.0006 & -2.34 $\pm$ 0.01 & 0.71 $\pm$ 0.02 & 2.57 $\pm$ 0.75 & 0.19 $\pm$ 0.05 \\
3C154 & 185.6/4.0 & 0.01 $\pm$ 0.0006 & -1.39 $\pm$ 0.04 & 0.83 $\pm$ 0.09 & 5.2 $\pm$ 5.28 & 0.18 $\pm$ 0.19 & &  0.0181 $\pm$ 0.0005 & -1.34 $\pm$ 0.02 & 0.94 $\pm$ 0.05 & 4.46 $\pm$ 1.79 & 0.18 $\pm$ 0.07 \\
3C154 & 185.6/4.0 & 0.0038 $\pm$ 0.0005 & 2.23 $\pm$ 0.07 & 1.14 $\pm$ 0.17 & 5.83 $\pm$ 6.56 & 0.11 $\pm$ 0.12 & &  0.0054 $\pm$ 0.0004 & 2.19 $\pm$ 0.05 & 1.57 $\pm$ 0.13 & 0.54 $\pm$ 8.69 & 0.01 $\pm$ 0.17 \\
3C167 & 207.3/1.2 & 0.0106 $\pm$ 0.0019 & 18.46 $\pm$ 0.12 & 1.49 $\pm$ 0.35 & 4.75 $\pm$ 17.95 & 0.32 $\pm$ 1.22 & &  0.009 $\pm$ 0.0018 & 17.77 $\pm$ 0.15 & 1.76 $\pm$ 0.49 & 4.59 $\pm$ 9.57 & 0.17 $\pm$ 0.36 \\
3C18 & 118.6/-52.7 & 0.0031 $\pm$ 0.0003 & -8.52 $\pm$ 0.11 & 2.64 $\pm$ 0.27 & 10.92 $\pm$ 14.88 & 0.38 $\pm$ 0.52 & &  0.006 $\pm$ 0.0003 & -8.34 $\pm$ 0.05 & 2.61 $\pm$ 0.14 & 9.2 $\pm$ 6.04 & 0.34 $\pm$ 0.23 \\
3C18 & 118.6/-52.7 & 0.0056 $\pm$ 0.0004 & -7.82 $\pm$ 0.02 & 0.67 $\pm$ 0.07 & 6.45 $\pm$ 3.7 & 0.1 $\pm$ 0.06 & &  0.0079 $\pm$ 0.0004 & -7.85 $\pm$ 0.01 & 0.6 $\pm$ 0.04 & 4.83 $\pm$ 1.6 & 0.05 $\pm$ 0.02 \\
3C207 & 213.0/30.1 & 0.015 $\pm$ 0.0002 & 4.55 $\pm$ 0.01 & 0.76 $\pm$ 0.01 & 2.94 $\pm$ 1.55 & 0.14 $\pm$ 0.08 & &  0.0266 $\pm$ 0.0002 & 4.55 $\pm$ 0.0 & 0.77 $\pm$ 0.01 & 2.48 $\pm$ 0.46 & 0.12 $\pm$ 0.02 \\
3C409 & 63.4/-6.1 & 0.0058 $\pm$ 0.0011 & 14.59 $\pm$ 0.27 & 1.68 $\pm$ 0.35 & 11.31 $\pm$ 8.53 & 0.47 $\pm$ 0.38 & &  0.0055 $\pm$ 0.0015 & 14.68 $\pm$ 0.33 & 1.52 $\pm$ 0.4 & 7.83 $\pm$ 11.73 & 0.16 $\pm$ 0.24 \\
3C409 & 63.4/-6.1 & 0.0204 $\pm$ 0.0025 & 15.4 $\pm$ 0.01 & 0.89 $\pm$ 0.05 & 3.18 $\pm$ 2.31 & 0.25 $\pm$ 0.18 & &  0.0275 $\pm$ 0.0032 & 15.42 $\pm$ 0.01 & 0.86 $\pm$ 0.04 & 0.62 $\pm$ 1.0 & 0.03 $\pm$ 0.06 \\
3C410 & 69.2/-3.8 & 0.0044 $\pm$ 0.0006 & 6.32 $\pm$ 0.04 & 1.89 $\pm$ 0.15 & 13.41 $\pm$ 11.22 & 0.47 $\pm$ 0.4 & &  0.0079 $\pm$ 0.0005 & 6.38 $\pm$ 0.02 & 2.32 $\pm$ 0.09 & 6.4 $\pm$ 5.25 & 0.28 $\pm$ 0.23 \\
3C410 & 69.2/-3.8 & 0.0089 $\pm$ 0.0006 & 6.21 $\pm$ 0.01 & 0.65 $\pm$ 0.04 & 8.46 $\pm$ 2.4 & 0.21 $\pm$ 0.06 & &  0.0193 $\pm$ 0.0005 & 6.26 $\pm$ 0.01 & 0.81 $\pm$ 0.02 & 3.81 $\pm$ 1.28 & 0.14 $\pm$ 0.05 \\
3C410 & 69.2/-3.8 & 0.0044 $\pm$ 0.0003 & 10.7 $\pm$ 0.03 & 0.71 $\pm$ 0.07 & 10.06 $\pm$ 5.89 & 0.13 $\pm$ 0.08 & &  0.0085 $\pm$ 0.0002 & 10.71 $\pm$ 0.02 & 0.81 $\pm$ 0.04 & 4.15 $\pm$ 3.09 & 0.07 $\pm$ 0.05 \\
3C410 & 69.2/-3.8 & 0.0054 $\pm$ 0.0002 & 11.67 $\pm$ 0.03 & 0.84 $\pm$ 0.07 & 4.83 $\pm$ 4.66 & 0.09 $\pm$ 0.09 & &  0.0115 $\pm$ 0.0002 & 11.68 $\pm$ 0.02 & 0.82 $\pm$ 0.03 & 2.93 $\pm$ 3.0 & 0.07 $\pm$ 0.07 \\
3C454.3 & 86.1/-38.2 & 0.0023 $\pm$ 0.0001 & -9.67 $\pm$ 0.03 & 1.6 $\pm$ 0.09 & 4.63 $\pm$ 12.36 & 0.07 $\pm$ 0.19 & &  0.0044 $\pm$ 0.0001 & -9.54 $\pm$ 0.01 & 1.25 $\pm$ 0.04 & 8.13 $\pm$ 6.06 & 0.1 $\pm$ 0.08 \\
3C75 & 170.3/-44.9 & 0.0071 $\pm$ 0.0005 & -10.36 $\pm$ 0.04 & 1.3 $\pm$ 0.12 & 3.45 $\pm$ 5.41 & 0.14 $\pm$ 0.21 & &  0.014 $\pm$ 0.0008 & -10.36 $\pm$ 0.03 & 1.22 $\pm$ 0.09 & 3.51 $\pm$ 1.56 & 0.14 $\pm$ 0.06 \\
4C13.67 & 43.5/9.2 & 0.0464 $\pm$ 0.0043 & 4.85 $\pm$ 0.05 & 1.1 $\pm$ 0.12 & 10.43 $\pm$ 2.77 & 2.28 $\pm$ 0.69 & &  0.0567 $\pm$ 0.0057 & 4.89 $\pm$ 0.05 & 1.12 $\pm$ 0.14 & 10.34 $\pm$ 2.13 & 1.55 $\pm$ 0.4 \\
4C22.12 & 188.1/0.0 & 0.0058 $\pm$ 0.001 & -2.84 $\pm$ 0.07 & 0.79 $\pm$ 0.19 & 6.54 $\pm$ 7.03 & 0.13 $\pm$ 0.14 & &  0.0102 $\pm$ 0.0011 & -2.73 $\pm$ 0.04 & 0.78 $\pm$ 0.12 & 6.72 $\pm$ 2.32 & 0.13 $\pm$ 0.05 \\
4C22.12 & 188.1/0.0 & 0.0172 $\pm$ 0.0012 & -1.78 $\pm$ 0.02 & 0.56 $\pm$ 0.05 & 5.07 $\pm$ 2.12 & 0.21 $\pm$ 0.09 & &  0.0354 $\pm$ 0.0013 & -1.78 $\pm$ 0.01 & 0.54 $\pm$ 0.03 & 3.78 $\pm$ 0.74 & 0.17 $\pm$ 0.03 \\
G196.6+0.2 & 196.6/0.2 & 0.0044 $\pm$ 0.0005 & 3.26 $\pm$ 0.11 & 1.94 $\pm$ 0.27 & 10.82 $\pm$ 12.22 & 0.4 $\pm$ 0.46 & &  0.0062 $\pm$ 0.0005 & 3.4 $\pm$ 0.09 & 2.38 $\pm$ 0.22 & 8.85 $\pm$ 8.6 & 0.31 $\pm$ 0.3 \\
G197.0+1.1 & 197.0/1.1 & 0.0126 $\pm$ 0.0005 & 4.83 $\pm$ 0.04 & 1.88 $\pm$ 0.09 & 6.94 $\pm$ 3.82 & 0.7 $\pm$ 0.39 & &  0.0191 $\pm$ 0.001 & 4.73 $\pm$ 0.04 & 1.65 $\pm$ 0.1 & 4.8 $\pm$ 2.17 & 0.36 $\pm$ 0.16 \\
G197.0+1.1 & 197.0/1.1 & 0.0059 $\pm$ 0.0009 & 7.46 $\pm$ 0.05 & 0.65 $\pm$ 0.11 & 0.31 $\pm$ 10.58 & 0.0 $\pm$ 0.17 & &  0.0078 $\pm$ 0.0015 & 7.34 $\pm$ 0.06 & 0.65 $\pm$ 0.14 & 1.61 $\pm$ 5.87 & 0.02 $\pm$ 0.07 \\
G197.0+1.1 & 197.0/1.1 & 0.0049 $\pm$ 0.0005 & 17.01 $\pm$ 0.12 & 2.47 $\pm$ 0.28 & 10.99 $\pm$ 9.9 & 0.57 $\pm$ 0.52 & &  0.0081 $\pm$ 0.0034 & 16.26 $\pm$ 0.17 & 0.91 $\pm$ 0.3 & 6.45 $\pm$ 2.95 & 0.11 $\pm$ 0.08 \\
G197.0+1.1 & 197.0/1.1 & 0.0052 $\pm$ 0.0015 & 17.59 $\pm$ 0.03 & 0.25 $\pm$ 0.08 & 9.87 $\pm$ 1.81 & 0.05 $\pm$ 0.03 & &  0.0127 $\pm$ 0.0013 & 17.38 $\pm$ 0.2 & 1.46 $\pm$ 0.35 & 5.46 $\pm$ 4.35 & 0.24 $\pm$ 0.2 \\
G197.0+1.1 & 197.0/1.1 & 0.0237 $\pm$ 0.001 & 32.01 $\pm$ 0.01 & 0.57 $\pm$ 0.03 & 4.75 $\pm$ 1.92 & 0.27 $\pm$ 0.11 & &  0.043 $\pm$ 0.0017 & 32.01 $\pm$ 0.01 & 0.54 $\pm$ 0.02 & 3.96 $\pm$ 1.04 & 0.22 $\pm$ 0.06 \\
P0428+20 & 176.8/-18.6 & 0.0014 $\pm$ 0.0002 & 3.6 $\pm$ 0.08 & 1.01 $\pm$ 0.19 & 13.48 $\pm$ 7.8 & 0.08 $\pm$ 0.05 & &  0.0029 $\pm$ 0.0003 & 3.54 $\pm$ 0.03 & 0.69 $\pm$ 0.08 & 4.45 $\pm$ 9.98 & 0.02 $\pm$ 0.05 \\
P0428+20 & 176.8/-18.6 & 0.0075 $\pm$ 0.0002 & 10.7 $\pm$ 0.02 & 1.09 $\pm$ 0.04 & 13.49 $\pm$ 3.45 & 0.47 $\pm$ 0.12 & &  0.0136 $\pm$ 0.0002 & 10.7 $\pm$ 0.01 & 1.1 $\pm$ 0.02 & 12.72 $\pm$ 1.62 & 0.45 $\pm$ 0.06 \\
T0526+24 & 181.4/-5.2 & 0.0172 $\pm$ 0.0073 & 7.55 $\pm$ 0.29 & 1.9 $\pm$ 1.13 & 13.7 $\pm$ 15.65 & 1.91 $\pm$ 2.59 & &  0.043 $\pm$ 0.0102 & 7.56 $\pm$ 0.14 & 2.43 $\pm$ 0.75 & 10.19 $\pm$ 7.5 & 2.52 $\pm$ 2.1 \\
T0629+10 & 201.5/0.5 & 0.0043 $\pm$ 0.0022 & 0.16 $\pm$ 0.0 & 0.65 $\pm$ 0.4 & 4.16 $\pm$ 2.97 & 0.05 $\pm$ 0.05 & &  0.0103 $\pm$ 0.0035 & 0.35 $\pm$ 0.0 & 1.18 $\pm$ 0.41 & 3.25 $\pm$ 1.93 & 0.09 $\pm$ 0.07 \\
T0629+10 & 201.5/0.5 & 0.0387 $\pm$ 0.0074 & 3.14 $\pm$ 0.13 & 1.1 $\pm$ 0.0 & 3.85 $\pm$ 0.47 & 0.7 $\pm$ 0.16 & &  0.0577 $\pm$ 0.0113 & 3.07 $\pm$ 0.14 & 1.1 $\pm$ 0.0 & 4.54 $\pm$ 0.55 & 0.68 $\pm$ 0.16 \\
T0629+10 & 201.5/0.5 & 0.0169 $\pm$ 0.0015 & 1.46 $\pm$ 0.07 & 1.39 $\pm$ 0.26 & 2.19 $\pm$ 2.11 & 0.22 $\pm$ 0.22 & &  0.0281 $\pm$ 0.0029 & 1.51 $\pm$ 0.08 & 1.23 $\pm$ 0.25 & 2.9 $\pm$ 0.91 & 0.24 $\pm$ 0.09 \\
T0629+10 & 201.5/0.5 & 0.1607 $\pm$ 0.0104 & 3.6 $\pm$ 0.01 & 0.61 $\pm$ 0.02 & 3.83 $\pm$ 0.35 & 1.59 $\pm$ 0.19 & &  0.2536 $\pm$ 0.0154 & 3.6 $\pm$ 0.01 & 0.65 $\pm$ 0.03 & 4.72 $\pm$ 0.64 & 1.84 $\pm$ 0.29 \\
T0629+10 & 201.5/0.5 & 0.0811 $\pm$ 0.002 & 4.62 $\pm$ 0.01 & 0.76 $\pm$ 0.03 & 6.43 $\pm$ 0.97 & 1.68 $\pm$ 0.27 & &  0.1553 $\pm$ 0.0037 & 4.61 $\pm$ 0.01 & 0.67 $\pm$ 0.03 & 6.62 $\pm$ 1.01 & 1.63 $\pm$ 0.26 \\
T0629+10 & 201.5/0.5 & 0.0747 $\pm$ 0.0018 & 6.09 $\pm$ 0.02 & 1.06 $\pm$ 0.05 & 5.44 $\pm$ 1.58 & 1.84 $\pm$ 0.54 & &  0.1165 $\pm$ 0.003 & 6.06 $\pm$ 0.02 & 1.17 $\pm$ 0.07 & 6.52 $\pm$ 1.67 & 2.1 $\pm$ 0.55 \\
T0629+10 & 201.5/0.5 & 0.0367 $\pm$ 0.0031 & 7.0 $\pm$ 0.02 & 0.49 $\pm$ 0.06 & 4.36 $\pm$ 0.68 & 0.33 $\pm$ 0.07 & &  0.0631 $\pm$ 0.0056 & 7.0 $\pm$ 0.02 & 0.5 $\pm$ 0.06 & 4.2 $\pm$ 0.3 & 0.31 $\pm$ 0.05 \\
T0629+10 & 201.5/0.5 & 0.0174 $\pm$ 0.0018 & 7.9 $\pm$ 0.05 & 0.83 $\pm$ 0.13 & 3.44 $\pm$ 1.67 & 0.21 $\pm$ 0.11 & &  0.0307 $\pm$ 0.003 & 7.91 $\pm$ 0.05 & 0.82 $\pm$ 0.13 & 3.65 $\pm$ 1.21 & 0.22 $\pm$ 0.08 \\

\hline
\end{tabular}
\end{sidewaystable}

\clearpage

\begin{table}
\begin{center}
\fontsize{7}{6}\selectfont
\caption{34 atomic sightlines}
\centering
\label{table:noh_vs_av}
\begin{tabular}{lccccccc}
\noalign{\smallskip} \hline \hline \noalign{\smallskip}
\shortstack{Sources\\ (Name)} & \shortstack{$l/b$\\$(^{o})$} & $\shortstack{\NHI\\(10$^{20}$cm$^{-2}$)}$ & \shortstack{\NHIthin\\(10$^{20}$cm$^{-2}$)} & 
\shortstack{$\sigma_{\tau}$(OH$_{1667}$)\\(10$^{-4}$)} & 
\shortstack{\NHm(upper limit)\tablenotemark{*}\\(10$^{20}$cm$^{-2}$)} & \shortstack{\t353\\(10$^{-6}$)} & \shortstack{\ebv\\(10$^{-2}$ mag)} \\
\hline
3C33 & 129.4/-49.3 & 3.25$\pm$0.0 & 3.2$\pm$0.1 & 12.14 & 0.6 & 2.16$\pm$0.07 & 3.54$\pm$0.42 \\
3C142.1 & 197.6/-14.5 & 25.11$\pm$2.6 & 19.6$\pm$0.8 & 10.55 & 0.52 & 21.53$\pm$0.72 & 21.71$\pm$0.81 \\
3C138 & 187.4/-11.3 & 22.9$\pm$1.1 & 19.9$\pm$0.3 & 5.02 & 0.25 & 21.63$\pm$0.81 & 17.47$\pm$0.59 \\
3C79 & 164.1/-34.5 & 10.86$\pm$1.2 & 9.8$\pm$0.8 & 37.03 & 1.84 & 9.23$\pm$0.37 & 12.67$\pm$0.78 \\
3C78 & 174.9/-44.5 & 11.69$\pm$0.5 & 10.3$\pm$0.2 & 13.25 & 0.66 & 12.45$\pm$0.63 & 14.64$\pm$1.07 \\
3C310 & 38.5/60.2 & 4.29$\pm$0.1 & 4.0$\pm$0.1 & 16.19 & 0.8 & 3.48$\pm$0.15 & 2.75$\pm$0.53 \\
3C315 & 39.4/58.3 & 5.48$\pm$0.4 & 4.7$\pm$0.0 & 12.96 & 0.64 & 3.98$\pm$0.09 & 5.63$\pm$0.26 \\
3C234 & 200.2/52.7 & 1.84$\pm$0.0 & 1.9$\pm$1.1 & 12.66 & 0.63 & 0.78$\pm$0.03 & 1.64$\pm$0.56 \\
3C236 & 190.1/54.0 & 1.38$\pm$0.0 & 1.3$\pm$1.2 & 10.72 & 0.53 & 0.7$\pm$0.03 & 2.04$\pm$0.49 \\
3C64 & 157.8/-48.2 & 7.29$\pm$0.2 & 6.9$\pm$0.8 & 33.12 & 1.65 & 6.55$\pm$0.35 & 9.01$\pm$0.36 \\
P0531+19 & 186.8/-7.1 & 27.33$\pm$0.7 & 25.4$\pm$0.3 & 6.37 & 0.32 & 20.75$\pm$0.62 & 20.54$\pm$1.3 \\
P0820+22 & 201.4/29.7 & 4.82$\pm$0.2 & 4.8$\pm$1.1 & 7.09 & 0.35 & 3.73$\pm$0.11 & 2.56$\pm$0.4 \\
3C192 & 197.9/26.4 & 4.56$\pm$0.1 & 4.5$\pm$0.1 & 20.66 & 1.03 & 3.38$\pm$0.08 & 4.05$\pm$0.53 \\
3C98 & 179.8/-31.0 & 12.7$\pm$0.5 & 11.3$\pm$1.3 & 12.26 & 0.61 & 13.72$\pm$0.41 & 17.73$\pm$1.02 \\
3C273 & 289.9/64.4 & 2.35$\pm$0.0 & 2.3$\pm$0.1 & 21.0 & 1.04 & 1.3$\pm$0.09 & 1.98$\pm$0.41 \\
DW0742+10 & 209.8/16.6 & 2.77$\pm$0.0 & 2.8$\pm$0.9 & 8.01 & 0.4 & 1.6$\pm$0.03 & 1.99$\pm$0.25 \\
3C172.0 & 191.2/13.4 & 8.89$\pm$0.2 & 8.6$\pm$1.1 & 13.02 & 0.65 & 5.66$\pm$0.08 & 4.7$\pm$0.52 \\
3C293 & 54.6/76.1 & 1.46$\pm$0.1 & 1.5$\pm$1.1 & 6.24 & 0.31 & 1.31$\pm$0.09 & 2.83$\pm$0.75 \\
3C120 & 190.4/-27.4 & 18.17$\pm$2.1 & 10.7$\pm$0.1 & 28.29 & 1.41 & 29.26$\pm$1.08 & 22.74$\pm$1.03 \\
CTA21 & 166.6/-33.6 & 10.97$\pm$0.4 & 10.0$\pm$0.8 & 27.35 & 1.36 & 10.39$\pm$0.44 & 13.3$\pm$0.96 \\
P1117+14 & 240.4/65.8 & 1.79$\pm$0.0 & 1.8$\pm$0.3 & 15.0 & 0.75 & 1.5$\pm$0.04 & 2.73$\pm$0.54 \\
3C264.0 & 237.0/73.6 & 1.97$\pm$0.0 & 2.0$\pm$0.4 & 6.29 & 0.31 & 1.64$\pm$0.08 & 2.88$\pm$0.34 \\
3C208.1 & 213.6/33.6 & 3.15$\pm$0.1 & 3.2$\pm$0.2 & 18.67 & 0.93 & 3.12$\pm$0.04 & 2.93$\pm$0.43 \\
3C208.0 & 213.7/33.2 & 3.41$\pm$0.1 & 3.5$\pm$0.2 & 19.69 & 0.98 & 3.38$\pm$0.07 & 4.37$\pm$0.47 \\
4C32.44 & 67.2/81.0 & 1.23$\pm$0.0 & 1.3$\pm$0.6 & 9.88 & 0.49 & 0.81$\pm$0.02 & 1.94$\pm$0.29 \\
3C272.1 & 280.6/74.7 & 2.82$\pm$0.0 & 2.8$\pm$0.3 & 10.24 & 0.51 & 1.73$\pm$0.28 & 2.43$\pm$0.31 \\
4C07.32 & 322.2/68.8 & 2.43$\pm$0.0 & 2.4$\pm$0.3 & 30.7 & 1.53 & 2.32$\pm$0.06 & 4.3$\pm$0.41 \\
3C245 & 233.1/56.3 & 2.39$\pm$0.0 & 2.4$\pm$0.2 & 11.36 & 0.56 & 2.22$\pm$0.06 & 2.71$\pm$0.36 \\
3C348 & 23.0/29.2 & 6.56$\pm$0.2 & 6.0$\pm$0.1 & 16.51 & 0.82 & 5.7$\pm$0.15 & 9.8$\pm$0.33 \\
3C286 & 56.5/80.7 & 2.33$\pm$0.1 & 2.4$\pm$2.7 & 7.13 & 0.35 & 0.81$\pm$0.05 & 2.75$\pm$0.74 \\
4C13.65 & 39.3/17.7 & 10.56$\pm$0.2 & 9.9$\pm$0.1 & 20.01 & 0.99 & 11.8$\pm$0.39 & 15.63$\pm$0.6 \\
3C190.0 & 207.6/21.8 & 3.21$\pm$0.0 & 3.4$\pm$0.9 & 17.57 & 0.87 & 2.41$\pm$0.03 & 1.96$\pm$0.42 \\
3C274.1 & 269.9/83.2 & 2.74$\pm$0.0 & 2.6$\pm$0.1 & 13.13 & 0.65 & 2.34$\pm$0.02 & 2.64$\pm$0.42 \\
3C298 & 352.2/60.7 & 2.39$\pm$0.4 & 2.6$\pm$0.1 & 10.69 & 0.53 & 1.3$\pm$0.07 & 1.97$\pm$0.39 \\
\noalign{\smallskip} \hline \noalign{\smallskip}

\multicolumn{8}{l}{\textsuperscript{*}\footnotesize{Estimated from OH(1667) 3$\sigma$ detection limits using T$_{\mathrm{ex}}$=3.5 K, FWHM=1 km/s and \NOH/\NHm=10$^{-7}$ (see Section \ref{sec:xoh}).}}

\end{tabular}
\end{center}
\end{table}

\end{document}